\documentclass[english,preprint,aps,prd,showpacs,superscriptaddress,nofootinbib,tightenlines]{revtex4}

\usepackage{amsfonts}
\usepackage{multirow}
\usepackage{mathrsfs}
\usepackage{graphicx}
\usepackage{amsmath}
\usepackage{amssymb}
\usepackage{bm}
\usepackage{bbm}
\usepackage{color}
\usepackage{float}
\usepackage{slashed}
\usepackage{caption}
\def\vslash{v\!\!\!\slash}
\def\zslash{z\!\!\!\slash}

\def\qslash{q\!\!\!\slash}

\def\varepsilonslash{\varepsilon\!\!\!\slash}

\newcommand{\nn}{\nonumber}

\newcommand{\beq}{\begin{equation}}
\newcommand{\eeq}{\end{equation}}
\newcommand{\bqa}{\begin{eqnarray}}
\newcommand{\eqa}{\end{eqnarray}}
\newcommand{\bseq}{\begin{subequations}}
\newcommand{\eseq}{\end{subequations}}

\begin{document}
%\preprint{}
%%%%%%%%%%%%%%%%%%%%%%%%%%%%%%%%%%%%%%%%%%%%%%%%%%%%%%%%%%%%%%%%%%%%%%%%%%%%%%
\title{\mbox{}\\[10pt]
$\bm{W}$ radiative decay to heavy-light mesons in HQET factorization through ${\cal O}(\alpha_s)$}
%%%%%%%%%%%%%%%%%%%%%%%%%%%%%%%%%%%%%%%%%%%%%%%%%%%%%%%%%%%%%%%%%%%%%%%%%%%%%%

\author{Saadi Ishaq~\footnote{saadi@ihep.ac.cn}}
\affiliation{Institute of High Energy Physics, Chinese Academy of Sciences, Beijing 100049, China}
\affiliation{School of Physical Sciences, University of Chinese Academy of Sciences,
Beijing 100049, China}

\author{Yu Jia~\footnote{jiay@ihep.ac.cn}}
\affiliation{Institute of High Energy Physics, Chinese Academy of
Sciences, Beijing 100049, China}
\affiliation{School of Physical Sciences, University of Chinese Academy of Sciences,
Beijing 100049, China}

\author{Xiaonu Xiong~\footnote{xnxiong@csu.edu.cn}}
\affiliation{School of Physics and Electronics, Central South University, Changsha 418003, China}
\affiliation{Institute for Advanced Simulation,
Institut f\"ur Kernphysik and J\"ulich Center for Hadron Physics,
Forschungszentrum J\"ulich, D-52425 J\"ulich, Germany}

\author{De-Shan Yang~\footnote{yangds@ucas.ac.cn}}
\affiliation{School of Physical Sciences, University of Chinese Academy of Sciences,
Beijing 100049, China}
\affiliation{Institute of High Energy Physics, Chinese Academy of
Sciences, Beijing 100049, China}

\date{\today}
%%%%%%%%%%%%%%%%%%%%%%%%%%%%%%%%%%%%%%%%%%%%%%%%%%%%%%%%%%%%%%%%%%%%%%%%%%%%%%
\begin{abstract}
Analogous to NRQCD factorization for heavy quarkonium exclusive production,
in this work we propose to employ the heavy-quark-effective-theory (HQET) factorization,
which has been predominantly applied to account for exclusive $B$ decays,
to study the exclusive production of the heavy-flavored mesons.
We take $W\to B(D_s)+\gamma$ as a prototype process. The validity of the
HQET factorization rests upon the presumed scale hierarchy: $m_W\sim m_b\gg \Lambda_{\rm QCD}$.
Through an explicit analysis at next-to-leading order in $\alpha_s$ yet at leading order in $1/m_b$,
we verify that the decay form factors can indeed
be expressed as the convolution between perturbatively calculable hard-scattering
kernel and the $B$ meson light-cone distribution amplitude (LCDA) defined in HQET.
It is observed that the factorization scale dependence becomes reduced after incorporating the NLO
perturbative correction.  An interesting future investigation is to identify and resum
large collinear logarithms of $m_W/m_b$ that arise ubiquitously in the fixed-order
expressions of the hard-scattering kernel in HQET factorization.
\end{abstract}

%\pacs{}

%%%%%%%%%%%%%%%%%%%%%%%%%%%%%%%%%%%%%%%%%%%%%%%%%%%%%%%%%%%%%%%%%%%%%%%%%%%%%%

\maketitle

\newpage{}
\section{Introduction}

Heavy flavor physics has been one of the major forefronts
of high energy physics in the past two decades,
whose primary goal is to trace the origin of the CP violation,
precisely pin down the Cabbibo-Kobayashi-Maskawa (CKM) matrix,
as well as search for possible footprints of new physics.
Much of effort in this subject concentrates on the detailed study of $B$ meson exclusive decays,
which contains very rich phenomenology.
To reliably describe uncountable decay channels, it is mandatory to develop a
systematic and thorough understanding towards the underlying hadronization dynamics.
By invoking the hierarchy $m_b\gg \Lambda_{\rm QCD}$,
the so-called QCD factorization (QCDF) approach has become the dominant theoretical arsenal
which derives from the first principle~\cite{Beneke:1999br}.
The key concept is to express the decay amplitude into the convolution of the perturbatively calculable, yet, process-dependent,
hard-scattering kernel and the nonperturbative, yet, universal, $B$ meson light-cone distribution amplitude (LCDA).
It is important to emphasize that this factorization framework applies in the heavy quark limit,
corroborated by the fact that the $b$ quark field in the LCDA is defined in the heavy quark effective theory (HQET)
rather than in full QCD. The simplest application of the QCD factorization is the radiative decay $B\to\gamma \ell \nu$ ~\cite{Korchemsky:1999qb,DescotesGenon:2002mw,Lunghi:2002ju}. This factorization picture has also been extended to more sophisticated processes,
{\it e.g.}, $B\to\gamma\gamma$~\cite{Bosch:2002bv}, $B\to M_1 M_2$~\cite{Beneke:2000ry,Beneke:2001ev}, and so on.

In contrast to exclusive $B$ decays, the hard exclusive production of heavy-flavored hadrons
receives much fewer attention in literature.
The main reason might be attributed to the highly suppressed production rate of
such processes in high energy collision experiments.
The $D_s$ meson exclusive production from $W$ decay, {\it e.g.}, $W^+\rightarrow D_{s}^+\gamma$, was among
the early exploration in this topics~\cite{Arnellos:1981gy,Keum:1993eb}, and an upper limit was placed by the
Fermilab \texttt{Tevatron} in late 90s~\cite{Abe:1998vm}.
Motivated by a gigantic number of $W^\pm$, $Z^0$ bosons produced at LHC, {\it i.e.}
about ${\cal O}(10^{11})$ with the projected $3000\;{\rm fb}^{-1}$  integrated luminosity,
a number of exclusive channels of $W$, $Z$ radiative decays into heavy-light mesons has recently been investigated
at leading order (LO) in $\alpha_s$~\cite{Grossmann:2015lea}.
The theoretical basis underneath \cite{Grossmann:2015lea} is the standard collinear factorization (or light-cone
factorization) for hard exclusive reactions~\cite{Lepage:1980fj,Chernyak:1983ej}.
By exploiting the hierarchy $m_{W,Z}\gg m_{b,c}\sim \Lambda_{\rm QCD}$, one expresses the amplitude as the convolution between the
hard-scattering kernel and the universal LCDAs of heavy mesons.
However, it is worth stressing that, the nonperturbative LCDAs encountered in this case are of
the standard Brodsky-Lepage type~\cite{Lepage:1980fj}, where the $b$ quark field is
defined in full QCD rather than in HQET, in sharp contrast with those arising in $B$ exclusive decays.
Unfortunately, our current constraints on the heavy meson QCD LCDAs
are rather limited, which severely obstructs the predictive power of the standard collinear factorization approach.
A further drawback of this conventional formalism is that, the LCDAs of heavy-flavor mesons cannot not
be genuinely nonperturbative, where the hard scale $m_{b,c}$ is still entangled with the hadronic scale
$\Lambda_{\rm QCD}$.
It is desirable if asymptotic freedom can be invoked to separate
some sort of short-distance effects from
the heavy meson QCD LCDAs.

Inspired by NRQCD factorization tailored for inclusive heavy quarkonium production~\cite{Bodwin:1994jh},
the {\it heavy-quark recombination} (HQR) mechanism~\cite{Braaten:2001bf}
was developed in the beginning of this century,
to supplement the single-parton fragmentation mechanism for the inclusive heavy-flavor hadron production
with contribution subleading in $1/p_\perp$.
The basic idea behind this mechanism is intuitively simple, after a hard scattering,
the heavy quark would have a significant chance to combine with a spectator quark which is {\it soft}
in its rest frame to form a heavy-light hadron. In the color-singlet channel,
where the inclusive and exclusive production of heavy hadrons practically make no difference at lowest order,
the HQR formalism only involves a single nonperturbative factor, which is proportional to the first inverse
moment of the $B$ meson LCDA defined in HQET, $1/\lambda_B$, which has been intensively investigated in the field
of $B$ meson decays. A notable success of this mechanism is to economically account for the
charm/anticharm hadron production asymmetry (leading particle effect) observed at numerous
Fermilab fixed-target experiments~\cite{Braaten:2001uu,Braaten:2002yt,Braaten:2003vy}.

The key idea behind HQR is to formalize the production rate of heavy-flavored hadrons
in the heavy quark limit, separating the dynamics of order $m_b$ or higher from
the hadronic effects at ${\cal O}(\Lambda_{\rm QCD})$,
but no longer distinguishing the hard-scattering scale specific to the process, say, $Q$, and $m_b$.
In this sense, one can tackle hard exclusive heavy hadron production in a fashion very much alike
the $B$ exclusive decay, in principle guaranteed by a factorization theorem valid to all orders in $\alpha_s$.
In this work, we choose not to stick to the old jargon HQR mechanism,
rather we decide to term this factorization approach as the {\it HQET factorization},
in close analogy with the NRQCD factorization
for heavy quarkonium production. To be specific, in this work we will take $W\to B(D_s)+\gamma$
as the prototype processes for heavy meson exclusive production.
To our knowledge, the HQR mechanism so far
is only illustrated at leading order (LO) in $\alpha_s$.
It is the very goal of this work to verify the validity of
HQET factorization through next-to-leading order (NLO) in $\alpha_s$,
which provides a much more informative revelation about nontrivial QCD dynamics.

The rest of the paper is structured as follows.
%-------------------------------
In Sec.~\ref{sec:form:factors:kinematics}, we first decompose the $W^+\to B^++\gamma$ amplitude
in terms of two Lorentz-invariant form factors,
and introduce some light-cone kinematical variables.
%-------------------------------
In Sec.~\ref{sec:rev:B:meson:LCDA}, we recap some essential features about the leading-twist
$B$ meson LCDA, which
enters the factorization theorem in many $B$ exclusive decay processes.
%-------------------------------
In Sec.~\ref{sec:HQET:factorization}, in analogy with the NRQCD factorization for exclusive heavy
quarkonium production, we propose the HQET factorization formalism
for exclusive heavy-light hadron production, presumably valid to all orders in $\alpha_s$.
%-------------------------------
In Sec.~\ref{sec:hard:scatter:kernel}, we first determine the hard-scattering kernel for
$W\to B+\gamma$ at tree level.
%-------------------------------
We then proceed to compute the ${\cal O}(\alpha_s)$ correction to this process.
We explicitly verify that the resulting soft IR pole can be properly factorized into the $B$ meson LCDA, which
establishes the correctness of HQET factorization at the first nontrivial order.
The IR-finite hard-scattering kernel at ${\cal O}(\alpha_s)$ is also deduced.
%-------------------------------
In Sec.~\ref{sec:phenomenology}, we present a comprehensive numerical prediction for the processes
$W^+\rightarrow B^+(D_s^+)+\gamma$, accurate through next-to-leading order in $\alpha_s$.
Assuming a simple exponential parametrization for $B$ meson LCDA defined in some initial scale,
we study its evolution behavior with different factorization scale.
It is found that including the ${\cal O}(\alpha_s)$ correction would significantly
reduce the LO decay rate.
%-------------------------------
Finally, we present a summary and an outlook in Sec.~\ref{sec:summary}.
%-------------------------------

%%%%%%%%%%%%%%%%%%%%%%%%%%%%%%%%%%%%%%%
\section{Decomposition of amplitude and light-cone kinematics}
%%%%%%%%%%%%%%%%%%%%%%%%%%%%%%%%%%%%%%%
\label{sec:form:factors:kinematics}
%%%%%%%%%%%%%%%%%%%%%%%%%%%%%%%%%%%%%%%

Let us first specify the kinematics for the process $W^+ \to B^+ +\gamma $.
The momenta of $W^+$, $B^+$ and $\gamma$ are designated by $Q$, $P$ and $q$, respectively,
with $Q=P+q$. They are subject to the on-shell conditions: $Q^2=m_W^2$, $P^2=m_B^2$, and $q^2=0$,
with $m_W$, $m_B$ standing for the masses of the $W$ and $B$ meson, respectively.
For future usage, we also introduce a dimensionless four velocity $v^\mu$ via
$P^{\mu}=m_{B} v^{\mu}$, obviously obeying $v^{2}=1$.
The polarization vectors of the $W$ and $\gamma$ are denoted by  $\varepsilon_{W}$ and
$\varepsilon_\gamma$.

In accordance with Lorentz invariance, the decay amplitude for
$W^+\to B^+\gamma$ can be decomposed as~\cite{Grossmann:2015lea}
%-------------------------------
\begin{align}
%-------------------------------
\mathcal {M}\left(W^+\rightarrow B^+\gamma\right) = {\frac {  { e_u e^2 } V_{ub}  } { 4\sqrt{2} \sin \theta_W}}\left(  \epsilon _ { \mu \nu \alpha \beta } \frac { P ^ { \mu } q ^ { \nu } \varepsilon _ {W} ^ { \alpha } \varepsilon _ { \gamma } ^ { * \beta } } { P \cdot q } F_V  +i \varepsilon _ { W } \cdot \varepsilon _ { \gamma } ^ { * }  F_A \right),
%-------------------------------
\label{Ampl:Lorentz:decomp}
%-------------------------------
\end{align}
%-------------------------------
where $e$ is the electric coupling constant, $e_{u}={+}\frac{2}{3}$ is the electric charge of the
$u$ quark, $\theta_W$ is the weak mixing angle, $V_{ub}$ denotes the CKM matrix element,
and $f_B$ signifies the $B$ meson decay constant.
$F_{V}(F_A)$ represent the Lorentz scalar form factors, which are affiliated with
vacuum-to-$B+\gamma$ matrix element mediated by the vector (axial-vector) weak current.
All the nontrivial QCD dynamics are encoded in these scalar form factors,
which are functions of $m_W$, $m_B$, and $\Lambda_{rm QCD}$.
Note that, since weak interaction violates the parity conservation, this exclusive process bears
two independent helicity amplitudes, which can be expressed via the linear
combination of these two form factors.

Squaring \eqref{Ampl:Lorentz:decomp}, averaging upon and summing over the
polarizations of $W$ and $\gamma$, one expresses the unpolarized decay rate in the $W$ rest frame as
%-------------------------------
\begin{align}
%-------------------------------
\Gamma\left(W^+\rightarrow B^+\gamma\right)={\frac{e_u^2 \pi\alpha^2}{48 \sin^2 \theta_W m_W^3}}\left|V_{ub}\right|^2\left(m_W^2-m_B^2\right)\left(\left|F_V\right|^2+\left|F_A\right|^2\right),
%-------------------------------
\label{eq:dcy_wdth}
%-------------------------------
\end{align}
%-------------------------------
with $\alpha\equiv e^2/4\pi$ denoting the QED fine structure constant.

To facilitate the discussion in the following sections, it is convenient to
set up the light-cone representation for the kinematics.
We first introduce two light-like reference vectors
$n_{\pm}^{\mu}\equiv\frac{1}{\sqrt{2}}(1,0,0,\mp 1)$,
which obey $n_\pm^2=0$ and $n_+\cdot n_-=1$.
In these light-cone basis, any four vector $a^{\mu}=(a^0,a^1,a^2,a^3)$ can
then be decomposed into
%-------------------------------
\beq
%-------------------------------
a^{\mu}=(n_{-}\cdot a)
n_{+}^{\mu}+ (n_{+}\cdot a)n_{-}^{\mu}+a^{\mu}_{\perp}\equiv
a^+ n_{+}^{\mu}+a^- n_{-}^{\mu} + a^{\mu}_{\perp},
%-------------------------------
\eeq
%-------------------------------
where ${a}^{\mu}_{\perp}=(0,a^{1},a^{2},0)$ is the transverse component of the four vector.
The scalar product of two four vectors then become
%-------------------------------
\beq
%-------------------------------
a\cdot b = a^+ b^- + a^- b^+ + a_\perp\cdot b_\perp.
%-------------------------------
\eeq
%-------------------------------

Were we interested in investigating the process $W\to B+\gamma$ within
the standard collinear factorization approach, it would be most natural to stay with
the rest frame of the $W$ boson, where the $B$ meson is energetic owing to $m_W\gg m_B$.
In this reference frame, we presume that the $B$ meson moves along the positive $\hat{z}$ axis,
while the photon flies in the opposite direction.
The light-cone representations for the momenta of the $B^+$ and $\gamma$ then become
%-------------------------------
\bseq
%-------------------------------
\begin{align}
%-------------------------------
P^\mu \Big|_{W\;{\rm rest\;frame}} &=(P^{+},P^{-},\boldsymbol{P}_{\bot})=\frac{1}{\sqrt{2}}\left(m_W,\frac{m_B^2}{m_W},\boldsymbol{0}_{\bot}\right),
%-------------------------------
\\
%-------------------------------
q^\mu \Big|_{W\;{\rm rest\;frame}} &=\left( q^{+},q^{-},\boldsymbol{q}_{\bot} \right)
=\frac{1}{\sqrt{2}}\left( 0,\frac{m_W^2-m_B^2}{m_W}, \boldsymbol{0}_{\bot}\right),
%-------------------------------
\label{photon:momentum:W:rest:frame}
%-------------------------------
\end{align}
%-------------------------------
\label{momenta:W:rest:frame}\eseq
%-------------------------------
where $P^-$ is suppressed by a factor $m_B^2/m_W^2$ relative to $P^+$.

Since the form factors $F_{V,A}$ themselves are Lorentz scalars, they can be computed in any reference frame.
In order to make the picture of HQET factorization more
transparent, as well as be closely connected with the exclusive $B$ decay channel $B\to \gamma (W^*\to) l\nu$,
it looks most natural to boost this process to the $B$ meson rest frame.
The corresponding momenta of $B$ and photon in the light-cone basis then read
%-------------------------------
\bseq
%-------------------------------
\begin{align}
%-------------------------------
P^\mu\Big|_{B\;{\rm rest\;frame}} &=(P^{+},P^{-},\boldsymbol{P}_{\bot})=\frac{1}{\sqrt{2}}\left(m_B,m_B,\boldsymbol{0}_{\bot}\right),
%-------------------------------
\\
%-------------------------------
q^\mu \Big|_{B\;{\rm rest\;frame}} &=(q^{+},q^{-}, \boldsymbol{q}_{\bot})
=\frac{1}{\sqrt{2}}\left(0, {m_W^2-m_B^2\over m_B}, \boldsymbol{0}_{\bot}\right).
%-------------------------------
\label{photon:momenta:B:rest:frame}
%-------------------------------
\end{align}
%-------------------------------
\label{momenta:B:rest:frame}\eseq
%-------------------------------
Note the photon becomes enormously energetic in this frame,
enhanced with respect its energy in the $W$ rest frame by a factor $m_W/m_B$.

%%%%%%%%%%%%%%%%%%%%%%%%%%%%%%%%%%%%%%%
\section{Review of $\bm{B}$ meson LCDA defined in HQET}
%%%%%%%%%%%%%%%%%%%%%%%%%%%%%%%%%%%%%%%
\label{sec:rev:B:meson:LCDA}
%%%%%%%%%%%%%%%%%%%%%%%%%%%%%%%%%%%%%%%

$B$ meson LCDA is a crucial nonperturbative entity that ubiquitously enters the
various exclusive $B$ decay processes. It is the same entity that also enters
the $B$ exclusive production process in HQET factorization framework.
In this section, we briefly recapitulate some of its essential features.

Let us consider the correlator composed of the light spectator quark and $b$ quark separated by light-like
distance, sandwiched between the vacuum and the $B$ meson with velocity $v^\mu$ (for simplicity, we actually will
work in the $B$ meson rest frame).
Its most general parametrization may be cast into the following form~\cite{Grozin:1996pq,Beneke:2000wa}~\footnote{Note
that we have intentionally put the $B$ meson in the bra rather than the ket,
since we are interested in $B$ production rather than decay in this work.}:
%-------------------------------
\beq
%-------------------------------
\langle B(v)|\bar
{u}_{\beta}(z)[z,0] h_{v,\alpha}(0)|0\rangle=\frac{i \hat
f_{B}m_{B}}{4}\left\{\left[2
\widetilde{\phi}_{B}^{+}(t)-\frac{\zslash}{t}
\Big(\widetilde{\phi}_{B}^{-}(t)-\widetilde{\phi}_{B}^{+}(t)\Big)\right]
\frac{1-\vslash}{2}\gamma_{5}\right\}_{\alpha\beta},
%-------------------------------
\label{eq:LCDA_twist_decomp}
%-------------------------------
\eeq
%-------------------------------
where $z^2 =0$, $t=v\cdot z$, and $\widetilde{\phi}_{B}^{\pm}$ are a pair of
nonperturbative functions of $t$. Here $u$ refers to the standard
QCD field for $u$ quark, and $h_v$ signifies the $\bar{b}$ quark field
with velocity label $v$ introduced in HQET.
$\alpha$, $\beta$ are spinor indices.
$\hat{f}_{B}$ signifies the $B$ meson decay constant defined in HQET as
\beq
%-------------------------------
\langle B(v)|\bar
{u}\gamma^\mu \gamma_5 h_{v}|0\rangle=i\hat{f}_{B}m_{B}v^\mu\,,
%-------------------------------
\label{eq:decayconstant_HQET}
%-------------------------------
\eeq
%-------------------------------
which can be converted from the QCD decay constant $f_{B}$ through
perturbative series~\cite{Eichten:1989zv,Neubert:1993mb}:
%-------------------------------
\begin{align}
%-------------------------------
f_B = \hat{f}_B (\mu_F) \left[ 1 - {\alpha_s C_F \over 4 \pi} \left(3\ln \frac { \mu_F}{m_b} + 2
\right) \right] +\mathcal{O}\left(\alpha_s^2\right).
%-------------------------------
\label{hat:f:matching:fB}
%-------------------------------
\end{align}
%-------------------------------
The light-like gauge link,
%-------------------------------
\beq
%-------------------------------
[z, 0]=\mathcal{P} \exp \left[-i g_{s} \int_{0}^{z} d
\xi^{\mu} A_\mu^{a}(\xi) t^{a}\right],
%-------------------------------
\eeq
%-------------------------------
has been inserted in \eqref{eq:LCDA_twist_decomp} to ensure gauge
invariance of the nonlocal quark bilinear. Here $t^{a}(a=1, \cdots, 8)$ signify the $SU(3)$ generators in
fundamental representation, and $\mathcal{P}$ indicates the path
ordering.

The phenomenologically relevant $B$ meson LCDAs are usually referred in momentum space,
which can be inferred by Fourier transforming the coordinate-space correlators in
\eqref{eq:LCDA_twist_decomp}~\cite{Grozin:1996pq, Braaten:2001bf}:
%-------------------------------
\beq
%-------------------------------
\Phi_{B}^{\pm}(\omega) \equiv i\hat{f}_{B}m_{B}\phi^\pm_{B}(\omega)=\frac{1}{v^{\pm}}\displaystyle\int
{dt\over 2\pi}\, e^{i\omega t}\langle B(v)|\bar
{u}(z)[z,0]\slashed n_\mp \gamma_5 h_{v}(0)|0\rangle\Big|_{z^{+},z^{\perp}=0}\,,
%-------------------------------
\label{eq:LCDA_HQET}
%-------------------------------
\eeq
%-------------------------------
where a pair of $B$ meson LCDAs are defined through
%-------------------------------
\begin{equation}
%-------------------------------
{\phi}^{\pm}_{B}(\omega)=\int_{0}^{\infty} {d t\over 2\pi} e^{i
\omega t} \widetilde\phi^{\pm}_{B}(t).
%-------------------------------
\end{equation}
%-------------------------------
Here $\omega$ indicates the ``+''-momentum carried by the spectator
quark in the $B$ rest frame, whose typical value is $\sim \Lambda_{\rm QCD}$.
By construction, $\phi^{\pm}_{B}(\omega)$ has nonvanishing support only when $\omega \in (0,\infty)$.
General principle constrains that as $\omega\to 0$,
$\phi^{+}_{B}(\omega)\propto \omega$, whereas $\phi^{-}_{B}(\omega)\propto 1$.

Note the light-cone correlator in \eqref{eq:LCDA_HQET} in general entails UV divergences,
and is subject to the renormalization involving the mixing among an infinite number of light-ray operators.
As a consequence, $\phi^{\pm}_{B}(\omega)$ become scale-dependent quantities.
Practically speaking, provided that we are only interested in the leading-power contribution
in $1/m_b$ expansion, we are justified to concentrate on the $B$ meson LCDA $\phi_B^{+}(\omega)$
and discard $\phi_B^{-}(\omega)$.
The evolution equation governing $\phi_B^{+}(\omega,\mu)$ was first correctly
written down by Lange and Neubert in 2003~\cite{Lange:2003ff}:
%-------------------------------
\begin{align}
%-------------------------------
\notag \frac { d } { d \ln \mu } \phi _ { B } ^ { + } ( \omega ,\mu ) =& - \frac { \alpha _ { s } C _ { F } } { 4 \pi } \int _ { 0 } ^ { \infty } d \omega ^ { \prime }\left\{\left( 4 \ln \frac { \mu} { \omega } - 2 \right) \delta \left( \omega - \omega ^ { \prime } \right) - 4 \omega \left[ \frac { \theta \left( \omega ^ { \prime } - \omega \right) } { \omega^{ \prime } \left( \omega ^ { \prime } - \omega \right) }\right.\right.
%-------------------------------
\\
%-------------------------------
& \left.\left.+ \frac { \theta \left( \omega - \omega ^ { \prime } \right) }
{ \omega \left( \omega - \omega ^ { \prime } \right) } \right] _ { + }\right\}\phi_{ B }^{+}
\left( \omega ^ { \prime } , \mu \right),
%-------------------------------
\label{LN:evolution:eq}
%-------------------------------
\end{align}
%-------------------------------
with $\mu$ the renormalization scale.
Note this renormalization group equation looks rather different from the
celebrated Efremov-Radyushkin-Brodsky-Lepage (ERBL) equation~\cite{earlyBL,Efremov:1979qk,earlyCZ},
which controls
the scale dependence of the meson LCDA defined in full QCD.

Once the profile of $\phi^{+}_{B}(\omega)$ is determined at some initial scale (say, $\mu_0 = 1$ GeV),
one can utilize \eqref{LN:evolution:eq} to obtain its profile at any other scale (usually $1\;{\rm GeV}\le \mu \le m_b$).
There exists some model-dependent studies on the properties of the
$B$ meson LCDAs using QCD sum rules~\cite{Braun:2003wx}.
The model-independent features of $\phi^{+}_{B}(\omega)$
have also been abstracted by applying the
operator-product-expansion technique~\cite{Lee:2005gza}.

It turns out that for hard heavy hadron exclusive production,
only $\phi^{+}_{B}(\omega)$ survives in the factorization theorem
in the heavy quark limit.
As was mentioned before, of central phenomenological relevance is
the first inverse moment of $\phi^{+}_{B}(\omega)$, usually referred to as $\lambda_B^{-1}$:
%-------------------------------
\beq
%-------------------------------
\lambda_B^{-1}(\mu) \equiv \int_0^\infty \frac{d\omega}{\omega} \phi^+_B(\omega,\mu).
%-------------------------------
\label{first:inv:moment}
%-------------------------------
\eeq
%-------------------------------
Intuitively, one expects $\lambda_B^{-1}\sim \Lambda_{\rm QCD}^{-1}$. Note this inverse moment
is also scale-dependent.

We also plan to study the NLO perturbative corrections to $W\to B+\gamma$. To this purpose, it is
also necessary to introduce the first and second logarithmic inverse moments by
%-------------------------------
\begin{align}
%-------------------------------
\lambda_B^{-1}\sigma_{B,n} (\mu) &\equiv
 -\int_0^\infty \frac{d\omega}{\omega} \ln^n\frac{\omega}{\mu} \phi^+_B(\omega,\mu),
\qquad n=1,2
%-------------------------------
\label{Log:first:inverse:moments}
%-------------------------------
\end{align}
%-------------------------------
which are scale dependent as well.

As a side remark, we finally mention that the positive Mellin moments of $\phi^+_B\big(\omega)$, that is,
$\int_0^\infty \! d\omega \,\omega^{N-1} \phi^+_B(\omega)$ ($N > 0$), are generally
UV divergent, thus become ill-defined. Usually one imposes a hard UV cutoff in the upper end of the integral
to regularize the UV divergence~\cite{Lee:2005gza}.

%%%%%%%%%%%%%%%%%%%%%%%%%%%%%%%%%%%%%%%
\section{HQET factorization for exclusive heavy-light meson production}
%%%%%%%%%%%%%%%%%%%%%%%%%%%%%%%%%%%%%%%
\label{sec:HQET:factorization}
%%%%%%%%%%%%%%%%%%%%%%%%%%%%%%%%%%%%%%%

In this section we shall state the exact form of HQET factorization for
heavy-light meson exclusive production.
Since this factorization picture naturally evolves from the previously developed heavy quark recombination mechanism,
especially for the color-singlet channel, it might be beneficial to
first elaborate on the underlying physical picture, by taking the $W\to B+\gamma$ channel as a concrete example.
Viewed in the rest frame of the $B^+$ meson, after a hard scattering the $\bar{b}$ quark
has a considerable chance to pick up the {\it soft} spectator $u$ quark to hadronize into the $B^+$ meson,
with the recombination probability proportional to the square of the nonperturbative
factor $\lambda_B^{-1}$ defined in \eqref{first:inv:moment}.
By emitting a highly energetic photon, the soft $u$ quark is necessarily transformed into a hard-collinear one,
which endorses the usage of the $B$ meson LCDA.

Let $k^\mu$ signify the momentum carried by the spectator $u$ quark inside the $B^+$ meson.
According to the recipe of the HQR mechanism~\cite{Beneke:2000wa},
one may obtain the $W^+\to B^+ +\gamma$ amplitude through making the following substitution
in the quark amplitude $W^+\to [\bar{b}(P-k) u(k)]^{(1)} + \gamma$~\footnote{This form
may look superficially different from the projectors adopted in \cite{Braaten:2001bf}.
Nevertheless, once the equation of motion and
Wandzura-Wilczek approximation are invoked, they can be proven to be equivalent~\cite{Beneke:2000wa}.}:
%-------------------------------
\begin{equation}
%-------------------------------
v_i (P\!-\!k) \bar{u}_j (k)\to {\delta_{ij}\over N_c} \frac{i \hat{f}_{B}
m_b}{4}\!\left\{\frac{1\!-\!\slashed v}{2}\!\left[\phi_{B}^{+}(\omega)
{\slashed n_{+}\over\sqrt{2}}\!+\!\phi_{B}^{-}(\omega) {\slashed
n_{-}\over \sqrt{2}}-\omega \phi_{B}^{-}(\omega) \gamma^{\mu}_{\perp}
\frac{\partial}{\partial k_{\perp\mu}}\right]\!\!
\gamma_{5}\right\}\Bigg|_{k=\omega v},
%-------------------------------
\label{HQR:spin:projector:B}
\end{equation}
%-------------------------------
where $i,j=1,2, \cdots, N_c$ are color indices and $N_c=3$.
The first Kronecker symbol serves the color-singlet projector.
After taking the momentum derivative on the quark amplitude, one then makes the substitution
$k\to \omega v$ and retain the most singular piece in the $\omega\to 0$ limit,
which is usually $\propto 1/\omega$.
Curiously, in the heavy quark limit the $\phi_{B}^{-}(\omega)$ turns out not to contribute,
and only $\phi_{B}^{+}(\omega)$ yields a nonvanishing contribution,
whose effect is simply encoded in the first inverse moment $\lambda_B^{-1}$.

A shortcut can be invoked to quickly reproduce the
heavy meson production amplitude from the HQR mechanism, with much less effort~\cite{Beneke:2000wa}.
Rather than consider $B^+$ meson production, one simply starts with the flavored qurakonium, $B_c$, production.
Assume the momenta are partitioned by two constitutes of the $B_c$ as
$p_c= \kappa P$ and $p_b = (1-\kappa) P$, where $P$ is the $B_c$ momentum
and $\kappa=m_c/(m_c+m_b)$.
One can then employs the familiar covariant spin projector for
quarkonium production at LO in velocity expansion:
%-------------------------------
\beq
%-------------------------------
v_i (p_b) \bar{u}_j (p_c)\to  \delta_{ij} {f_{B_c}\over 12}
 \left(\slashed P - m_{B_c}\right) \gamma_5.
%-------------------------------
\label{HQR:spin:projector:Bc}
%-------------------------------
\eeq
%-------------------------------
Consequently, the amplitude for $B^+$ production through the $[\bar{b} u]({}^1S_0^{(1)})$
channel can be deduced by taking the $\kappa \to 0$ limit of that for $B_c$
production and replacing $f_{B_c}/(4\kappa)\to {\hat{f}_B}/{(4\lambda_B)}$.
In general, many Feynaman diagrams do not contribute for their lack of the
$1/\kappa$ singularity. Notice that, this shortcut has been utilized to ascertain
the NLO perturbative corrections to the $B$ electromagnetic form factor and $W\to B+\gamma$
from their $B_c$ counterparts in NRQCD factorization~\cite{Jia:2010fw,Feng:2019meh}.

To the best of our knowledge, the HQR mechanism thus far has not yet been extended to NLO in $\alpha_s$.
It is not straightforward to achieve this goal from the projector approach specified in
\eqref{HQR:spin:projector:B}, or from the shortcut of extracting the hard-scattering kernel from $B_c$ production,
as described in \eqref{HQR:spin:projector:Bc}.
One reason might be that $\phi_B^+$ develops UV divergence at NLO
in $\alpha_s$, therefore the hard-scattering kernel also acquires an explicit factorization scale dependence.
This is in sharp contrast with the one-loop correction to exclusive quarkonium production process,
since the local NRQCD bilinear operator is one-loop UV finite.
It is not {\it a priori} obvious why the aforementioned projector approach
can lead to the IR-finite hard-scattering kernel once beyond $\alpha_s$.

Since we treat $m_W$ as the same order as $m_b$, the argument that leads to the factorization theorem for
$B\to\gamma l\nu$ can be transplanted to $W\to B+\gamma$ without modification. We thus propose a
factorization theorem for $W\to B+\gamma$, which is valid in the heavy quark limit, albeit
to all orders in $\alpha_s$:
%-------------------------------
\begin{align}
%-------------------------------
{\mathcal M}(W^+\to B^+\gamma) = \hat{f}_B(\mu_F) \int_0^\infty \!\! d\omega\, T(\omega, m_b, \mu_F)
\phi^+_B(\omega,\mu_F)+\mathcal{O}\left(m_b^{-1}\right),
%-------------------------------
\label{HQET:factorization:theorem}
%-------------------------------
\end{align}
%-------------------------------
where $T(\omega,m_b, \mu_F)$ is referred to as the hard-scattering kernel, which can be computed in  perturbation theory.
Note the hard-scattering kernel is explicitly dependent on $m_W$ and $m_b$ as well as the factorization scale $\mu_F$.
Note only the leading-twist $B$ meson LCDA, $\phi^+_B(\omega)$, explicitly enters the formula.
The $\mu_F$ dependence of the hard-scattering kernel should counteract that of the $\hat{f}_B$ and $\phi_B^+$
so that the physical amplitude gets insensitive to the artificial scale $\mu_F$.
An important characteristic of this process is $T(\omega)\propto 1/\omega$,
so the convolution integral is UV finite and well-defined.

Equation~\eqref{HQET:factorization:theorem} lays down the foundation for the HQET factorization.
In some sense, \eqref{HQET:factorization:theorem} offers a systematic realization of
the HQR mechanism. The virtue of this factorization framework is to allow us to systematically investigate the
higher-order perturbative corrections for the hard-scattering kernel,
to ensure its IR finiteness.

%%%%%%%%%%%%%%%%%%%%%%%%%%%%%%%%%%%%%%%
\section{Form factors for $W\to B+\gamma$ through NLO in $\alpha_s$}
%%%%%%%%%%%%%%%%%%%%%%%%%%%%%%%%%%%%%%%
\label{sec:hard:scatter:kernel}
%%%%%%%%%%%%%%%%%%%%%%%%%%%%%%%%%%%%%%%

This section reports the central results of this work, where the hard-scattering kernel is computed
through NLO in $\alpha_s$.
Rather than employ the projector approach given in \eqref{HQR:spin:projector:Bc},
we choose to employ the perturbative matching method to determine the hard-scattering kernel.
The calculation presented in this section closely follows the analogous NLO calculation
for $B\to\gamma l\nu$~\cite{DescotesGenon:2002mw}.
Unless otherwise specified, the calculation of the form factors $F_{V,A}$ is performed
in the rest frame of the $B$ meson.

The hard-scattering kernel is insensitive to the long-distance physics.
For the sake of extracting it using perturbation theory, it is legitimate to replace the physical $B^+$ meson in
\eqref{HQET:factorization:theorem} with a fictitious $B^+$ meson
composed of a pair of free quarks $[\bar{b}(P-k) u(k)]$.
One can act the following projector on the quark amplitude:
%-------------------------------
\beq
%-------------------------------
v_i (P-k) \bar{u}_j (k)\to  {\delta_{ij} \over N_c}
 \frac{1-\slashed v}{4} \gamma_5,
%-------------------------------
\label{eq:projector:bbaru}
%-------------------------------
\eeq
%-------------------------------
just similar to the projector (\ref{HQR:spin:projector:Bc}) which
guarantees the fictitious ``$B^+$ meson'' being the color and spin singlet.
The momentum of the spectator $u$ quark is assumed to be soft, {\it i.e.}, which
scales as $k^\mu\sim \Lambda_{\rm QCD}$.
The LCDA in \eqref{eq:LCDA_HQET} for such a fictitious $B^+$ meson
then becomes
%-------------------------------
\beq
%-------------------------------
\Phi_{[\bar b u]}^{\pm}(\omega)=\frac{1}{v^{\pm}}\displaystyle\int
{dt\over 2\pi}\, e^{i\omega t}\langle [\bar b u](P)|\bar
{u}(z)[z,0]\slashed n_\mp \gamma_5 h_{v}(0)|0\rangle\Big|_{z^{+},z^{\perp}=0}.
%-------------------------------
\label{eq:LCDA_bbaru}
%-------------------------------
\eeq
%-------------------------------

Both of ingredients in \eqref{HQET:factorization:theorem}
can then be expanded in perturbation theory:
%-------------------------------
\beq
%-------------------------------
\Phi_{[\bar b u]}^+ = \Phi_{[\bar b u]}^{+(0)} + \Phi_{[\bar b u]}^{+(1)}+{\cal O}(\alpha_s^2),\qquad
\qquad
T = T^{(0)} + T^{(1)}+{\cal O}(\alpha_s^2),
%-------------------------------
\label{Expand:Phi:T}
%-------------------------------
\eeq
%-------------------------------
with the superscript indicating the powers of $\alpha_s$.
At LO, the LCDAs for the fictitious $B^+$ meson \eqref{eq:LCDA_bbaru} looks exceedingly simple
%-------------------------------
\beq
%-------------------------------
\Phi_{[\bar b u]}^{\pm (0)}(\omega)=\frac{1}{v^{\pm}}\delta\left(k^{+}/v^{+}-\omega\right)\text{Tr}\Big[ \frac{1-\slashed v}{4} \gamma_5 \slashed n_\mp\gamma_5\Big]=\delta\left(k^{+}/v^{+}-\omega\right).
%-------------------------------
\label{eq:LCDA_tree}
%-------------------------------
\eeq
%-------------------------------

The QCD amplitude on the left side of \eqref{HQET:factorization:theorem} can also be computed
perturbatively. Through NLO in $\alpha_s$, it reads
%-------------------------------
\beq
%-------------------------------
\mathcal{M} =\mathcal{M}^{(0)}+ \mathcal{M}^{(1)}+{\cal O}(\alpha_s^2),
%-------------------------------
\label{Expanding:fact:ampl:NLO:alphas}
%-------------------------------
\eeq
%-------------------------------
where
%-------------------------------
\bseq
%-------------------------------
\bqa
%-------------------------------
\mathcal{M}^{(0)} &=& \Phi_{[\bar b u]}^{+(0)}\otimes T^{(0)},
%-------------------------------
\label{Def:M0:LO:alphas}
%-------------------------------
\\
%-------------------------------
\mathcal{M}^{(1)} &=& \Phi_{[\bar b u]}^{+(0)}\otimes T^{(1)} + \Phi_{[\bar b u]}^{+(1)}\otimes T^{(0)},
%-------------------------------
\label{Def:M1:NLO:alphas}
%-------------------------------
\eqa
%-------------------------------
\label{M0:M1:pert:ampl}
%-------------------------------
\eseq
%-------------------------------
with $\otimes$ signifying the convolution integral in $\omega$.
Since both ${\cal M}$ and $\Phi_{[\bar b u]}^+$ are perturbatively calculable
for such a fictitious $B^+$ meson, one can solve \eqref{M0:M1:pert:ampl} iteratively,
to ascertain $T$ order by order in $\alpha_s$.

%%%%%%%%%%%%%%%%%%%%%%%%%%%%%%%%%%%%%%%
\subsection{Tree level}
%%%%%%%%%%%%%%%%%%%%%%%%%%%%%%%%%%%%%%%
\label{sec:tree:level:hard:kernel}
%%%%%%%%%%%%%%%%%%%%%%%%%%%%%%%%%%%%%%%

\begin{figure}
\centering
\includegraphics[clip,width=0.8\textwidth]{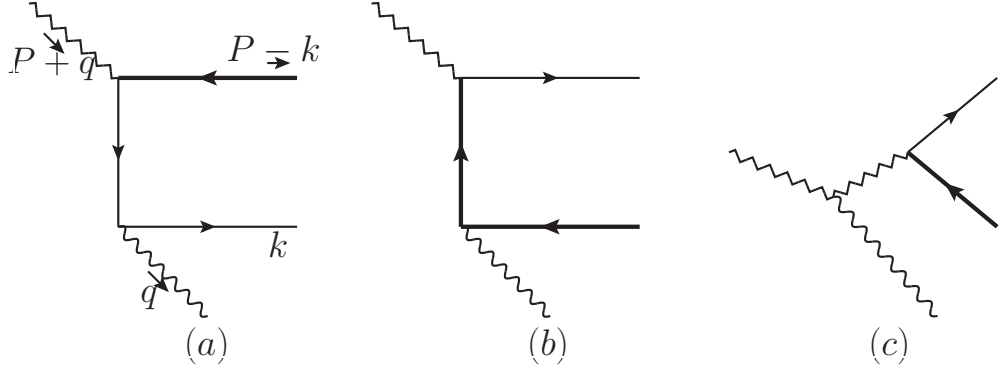}
\caption{The Feynman diagrams for $W^{+}\rightarrow [\bar{b} u]+\gamma$ at tree level.
The bold line represents the $\bar b$ quark.
}
\label{fig:tree}
\end{figure}

As depicted in Fig.~\ref{fig:tree}, there arise three electroweak diagrams at lowest order
that contribute to the quark-level process $W\to [\bar{b}(P-k) u(k)]^{(1)}+\gamma$.
Recall the momentum of the outgoing photon in \eqref{photon:momenta:B:rest:frame}
scales as $\left(q^+,q^-,|{\bf q}_\perp|\right)\sim(0,m_b, 0)$.
The $u$ propagator in Fig.~\ref{fig:tree}$a)$ becomes hard-collinear, and contribute a
$1/q\cdot k\sim 1/k^+q^-$ singularity to the amplitude. One can readily convince oneself that the other two diagrams,
the one with photon emitted from the $\bar{b}$ quark (Fig.~\ref{fig:tree}$b$) and the one emitted via a $WW\gamma$ vertex
(Fig.~\ref{fig:tree}$c$) do not possess such a $1/k^+$ enhancement, thus can be safely dropped.

Therefore, Fig.~\ref{fig:tree}$a)$ yields the following tree-level QCD amplitude
in the heavy quark limit:
%-------------------------------
\bqa
%-------------------------------
&&\mathcal{M}^{(0)}(W^{+}\rightarrow [\bar{b} u] + \gamma)
=\frac{e V_{ub} }{2\sqrt{2}\sin \theta_W} \left\langle[\bar b u] (P)\gamma (q,\varepsilon_\gamma)\left|\overline{u}\slashed\varepsilon_W(1-\gamma_{5})b\right|0\right\rangle\nonumber\\
%&=&\frac{e_{u}f_B}{2q^{-}k^{+}}\overline{u}(k)\epsilonslash^{\ast}\qslash\epsilonslash_W(1-\gamma_{5})v(P-k).\label{eq:M_LO}%T1
&\approx&{\frac{e_{u} e^2 V_{ub} }{4\sqrt{2}\sin\theta_W q^-
k^+}}\text{Tr}\Big[ \frac{1-\slashed v}{4} \gamma_5 \varepsilonslash_\gamma^{\ast}\qslash\varepsilonslash_W(1-\gamma_{5})\Big]
%-------------------------------
\nn\\
%-------------------------------
&=&{\frac{e_{u} e^2 V_{ub} }{4\sqrt{2}\sin\theta_W
}}\left(  -i\frac{\epsilon _ { \mu \nu \alpha \beta } v ^ { \mu } n_- ^ { \nu } \varepsilon _ {W} ^ { \alpha } \varepsilon _ { \gamma } ^ { * \beta } }{v^+} + \varepsilon _ { W } \cdot \varepsilon _ { \gamma } ^ { * }   \right)\int_0^\infty {d\omega\over\omega} \delta\left(k^+/v^+-\omega\right).
%-------------------------------
\label{eq:M_LO}
%-------------------------------
\eqa
%-------------------------------

With the aid of \eqref{eq:LCDA_tree} and \eqref{Def:M0:LO:alphas},
it is then straightforward to solve for
the tree-level hard-scattering kernel:
%-------------------------------
\begin{equation}
%-------------------------------
T^{(0)}(\omega)={\frac{e_{u} e^2 V_{ub} }{4\sqrt{2}\sin\theta_W
}}\left(  -i\frac{\epsilon _ { \mu \nu \alpha \beta } P^ { \mu }q ^ { \nu } \varepsilon _ {W} ^ { \alpha } \varepsilon _ { \gamma } ^ { * \beta } }{P\cdot q} + \varepsilon _ { W } \cdot \varepsilon _ { \gamma } ^ { * }   \right)\frac{1}{\omega}\,.
%-------------------------------
\label{eq:T_0}
%-------------------------------
\end{equation}
%-------------------------------
Comparing with the Lorentz decomposition specified in \eqref{Ampl:Lorentz:decomp},
one can deduce the final expressions for the vector/axial-vector form factors at tree level:
%-------------------------------
\begin{align}
%-------------------------------
F_V^{(0)}=F_A^{(0)}&={\hat{f}_B m_B} \int_0^\infty \frac{d\omega}{\omega} \phi_B^+\big(\omega\big)
={\frac{\hat{f}_B m_B}{\lambda_B}}.
%-------------------------------
\label{FVA:LO:expression}
%-------------------------------
\end{align}
%-------------------------------

We stress that the $\Phi_{[\bar{b}u]}^-(\omega)$ indeed does not enter the factorization formula.
Starting from \eqref{HQR:spin:projector:B}, inspecting the spinor structure of \eqref{eq:M_LO},
one can prove the $\Phi_{[\bar{b}u]}^-$-dependent terms vanish due to some specific identities such
as $n_-^2=0$ and $\gamma_\perp^\mu\varepsilonslash_\gamma^{\ast}\gamma_{\perp\mu}=(4-d)\varepsilonslash_\gamma^{\ast}$,
with $d=4$ signifying the spacetime dimension.

%%%%%%%%%%%%%%%%%%%%%%%%%%%%%%%%%%%%%%%
\subsection{One-loop level}
%%%%%%%%%%%%%%%%%%%%%%%%%%%%%%%%%%%%%%%
\label{sec:NLO:alphas:hard:kernel}
%%%%%%%%%%%%%%%%%%%%%%%%%%%%%%%%%%%%%%%

\begin{figure}
\centering
\includegraphics[clip,width=0.85\textwidth]{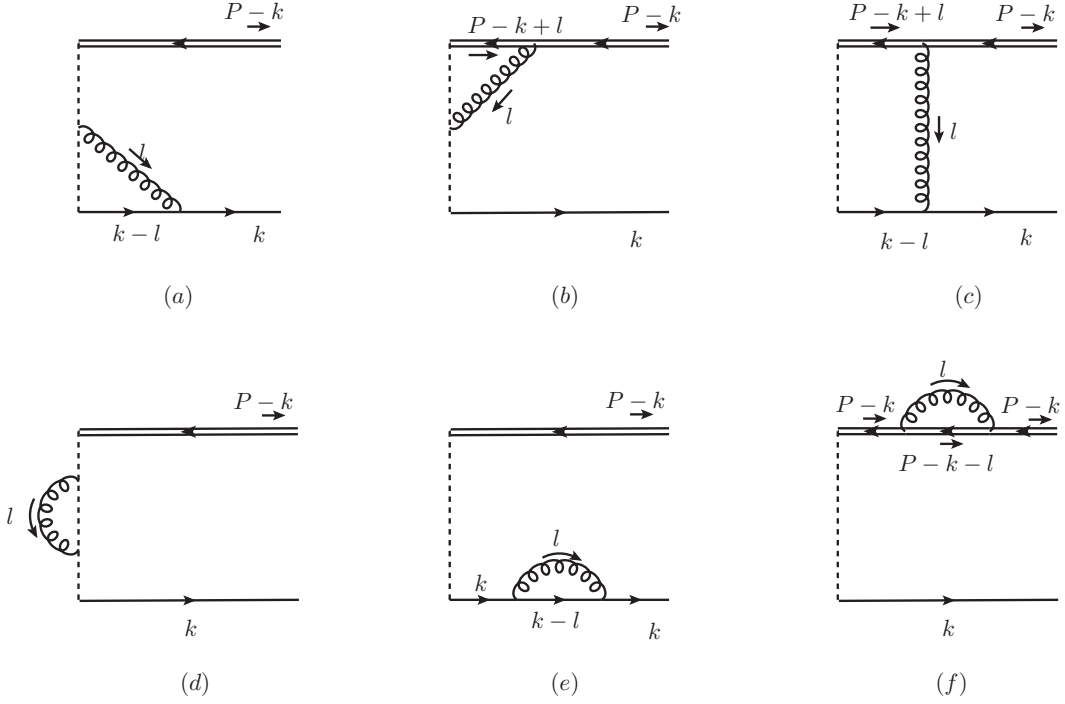}
\caption{One-loop QCD correction to LCDA for a fictitious $B$ meson.
The double line represents the $\bar{b}$ field in HQET, dashed line represents the gauge link.}
%It is unnecessary to calculate the subdiagrams $c)$ and $e)$.}
\label{Fig:One:loop:LCDA}
\end{figure}

\begin{figure}
\centering
\includegraphics[clip,width=0.8\textwidth]{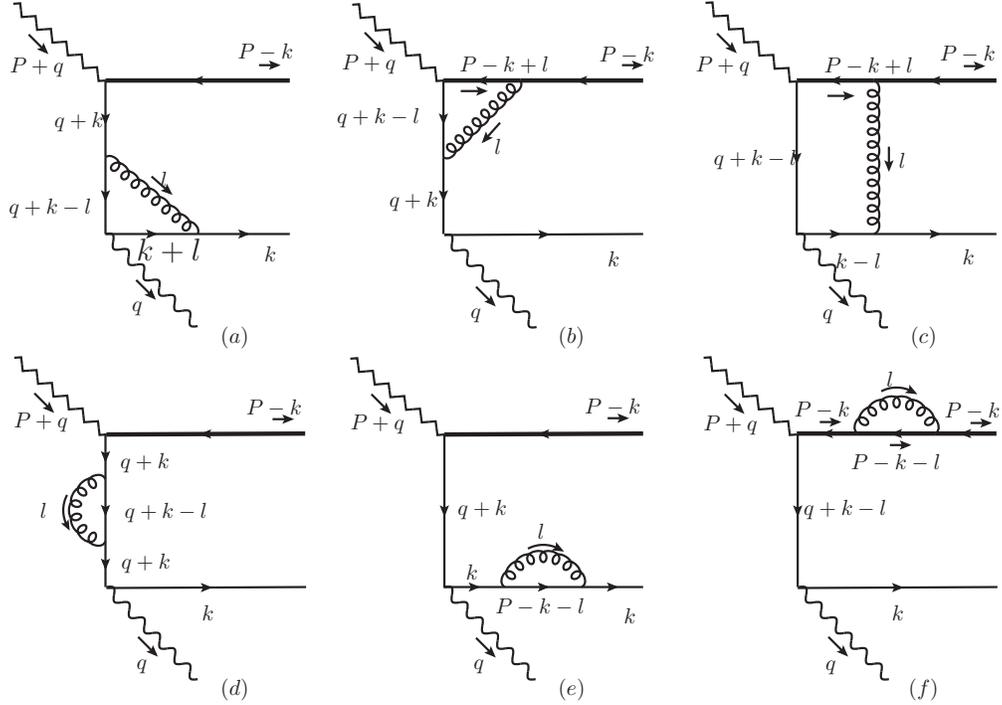}
\caption{One-loop QCD correction to the amplitude for $W^{+}\rightarrow
[\bar{b}u]^{(1)}+\gamma$. We just retain those diagrams with photon emitted from the spectator $u$ quark,
which yield leading contribution in the heavy quark limit.}
\label{Fig:One:loop:Amplitude}
\end{figure}

In this subsection, we proceed to extract the hard-scattering kernel through order-$\alpha_s$.
Following the ansatz given in \eqref{Def:M1:NLO:alphas},
the one-loop hard-scattering kernel $T^{(1)}$ can be extracted via
%-------------------------------
\beq
%-------------------------------
\Phi^{(0)}\otimes T^{(1)}=\mathcal{M}^{(1)}-\Phi^{(1)}\otimes T^{(0)},
%-------------------------------
\label{Extracting:T1}
%-------------------------------
\eeq
%-------------------------------
The one-loop diagrams for $\Phi^{(1)}$ and $\mathcal{M}^{(1)}$ are depicted in Fig.~\ref{Fig:One:loop:LCDA}
and Fig.~\ref{Fig:One:loop:Amplitude}, respectively.

By the general principle of effective field theory, both entities in the right-hand side of \eqref{Extracting:T1}
must possess the {\it identical} IR divergences, so that upon subtraction, $T^{(1)}$ must be infrared finite.
It appears instructive to compute the difference in \eqref{Extracting:T1}
on a diagram-by-diagram basis.

We employ dimensional regularization (with spacetime dimensions
$d=4-2 \epsilon$) to regularize UV divergences.
We treat $\gamma_5$ in the naive dimensional regularization (NDR) scheme
in which $\gamma_5$ anti-commutes with $\gamma^\mu$ ($\mu=0,1,...,d-1$).
We affiliate the 't Hooft unit mass $\mu_R$ when calculating
the QCD amplitude $\mathcal{M}^{(1)}$.
We also affiliate a different 't Hooft unit mass $\mu_F$
in computing $\Phi^{(1)}$.
When evaluating loop integrals, we have redefined the 't Hooft unit mass through
$\mu^2\to \mu^2 {e^{-\gamma_E}\over 4\pi}$,  which expedites the renormalization of
the LCDA according to the $\overline{\rm MS}$ subtraction scheme.

Moreover, a nonzero mass $m_u$ is retained for the spectator $u$ quark to regularize the
mass (collinear) singularity. For simplicity we adopt the Feynman gauge.
The Feynman rules for einkonal vertex and propagator
are given by $-ig_sT^a n_+^\mu$ and $1/p^+$, respectively,
with $p$ denoting the momentum flowing into the gauge link~\cite{Collins:2011zzd}.

The one-loop contributions to the perturbative ``$B^+$ meson'' LCDA,
as indicated in Fig.~\ref{Fig:One:loop:LCDA}$a)$, $b)$ and $e)$
can also be extracted from the {\it soft} loop region of
the electromagnetic vertex correction, weak vertex correction,
and light quark propagator correction for their QCD counterparts as shown
in Fig.~\ref{Fig:One:loop:Amplitude}.
The contributions from these one-loop bare diagrams
turn out to be~\cite{DescotesGenon:2002mw}
%-------------------------------
\bseq
%-------------------------------
%-------------------------------
\begin{align}
%-------------------------------
\Phi_{+\mathrm{em}\!}^{(1)}\otimes T^{(0)}= &
 \frac{\alpha_s C_F}{4\pi}\left(\frac{2}{\epsilon}\!-\!4\ln\frac{m_u}{\mu_{F}}\!+\!4\right)\mathcal{M}^{(0)},
%-------------------------------
\\
%-------------------------------
 \Phi_{+\mathrm{wk}\!}^{(1)}\otimes T^{(0)}=&
\frac{\alpha_s C_F}{4\pi}\left(\frac{1}{\epsilon^{2}}\!-\!\frac{2}{\epsilon}\ln\frac{k^{+}}{v^+\mu_{F}}
+2\ln^{2}\frac{k^{+}}{v^+\mu_{F}}\!+\!\frac{3\pi^{2}}{4}\right)\mathcal{M}^{(0)},
%-------------------------------
\label{Weak:vertex:LCDA:one:loop}
%-------------------------------
\\
%-------------------------------
\Phi^{(1)}_{+\delta Z_u}\!\otimes T^{(0)} =&
\frac{1}{2}\delta Z_u(\mu_F)\mathcal{M}^{(0)},
%-------------------------------
\end{align}
%-------------------------------
\eseq
%-------------------------------
where $\delta Z_u$ is the standard one-loop quark wave function renormalization constant in QCD.
Note that the occurrence of the double UV pole in \eqref{Weak:vertex:LCDA:one:loop} is
a peculiar trait of the HQET LCDA, from which one can infer the cusp anomalous dimension.

The gauge link self-energy diagram Fig.~\ref{Fig:One:loop:LCDA}$d)$ does not contribute to $\Phi_{+}^{(1)}\otimes T^{(0)}$ because its contribution is proportional to $n^2_+=0$.

To extract $T^{(1)}(\omega)$ following the recipe outlined in \eqref{Extracting:T1},
we need also calculate $\mathcal{M}^{(1)}$ indicated by those one-loop QCD diagrams in Fig.~\ref{Fig:One:loop:Amplitude}.
At leading power in $1/m_b$, we are only interested in retaining the contribution to
$\mathcal{M}^{(1)}$ of order $\Lambda_{\rm QCD}^{-1}$.
It is for this reason that we have excluded those diagrams where the photon is emitted
off the $\bar b$ quark or the $W^+$ boson.

It is straightforward to calculate the electromagnetic vertex correction, weak vertex correction a
nd internal quark self-energy QCD diagrams,
which have been depicted in Fig.~\ref{Fig:One:loop:Amplitude} $a)$, $b)$ and $d)$, respectively
(We have also tacitly included those quark mass counterterm diagrams in order to obtain UV-finite
results).
All three diagrams possess the same Lorentz structure as $\mathcal{M}^{(0)}$,
hence the corresponding contributions to $T^{(1)}(\omega)$ can be readily extracted by
subtracting the respective contributions of type $\Phi_{+}^{(1)}\otimes T^{(0)}$ in Fig.~\ref{Fig:One:loop:LCDA}.
We then obtain~\footnote{Note the matching procedure specified in \eqref{Extracting:T1} involves
the difference between two {\it renormalized} quantities. Including the quark wave-function and mass renormalization,
the one-loop QCD amplitude becomes UV finite and free from $\mu_R$ dependence; for the HQET LCDA,
one tacitly utilizes the $\overline{MS}$ subtraction scheme to render it finite. For simplicity,
in the following we will drop the UV poles in the hard-scattering
kernel associated with each individual Feynaman diagram.}
%-------------------------------
\bseq
%-------------------------------
\begin{align}
%-------------------------------
T_{\mathrm{em}}^{(1)} (\omega)= &\frac{\alpha_s C_F}{4\pi}T^{(0)}(\omega)\left(\ln\frac{2q^{-}v^{+} \omega}{\mu_F^2}+2\ln\frac{\mu_R}{\mu_F}-4-i\pi\right),
%-------------------------------
\\
%-------------------------------
\notag T_{\mathrm{wk}}^{(1)} (\omega)  =&  \frac{\alpha _s
C_F}{4\pi}T^{(0)}(\omega)\left\{\ln \frac{2 q^{-}
v^{+}\omega}{\left(1-r\right) m_W^2}
\left[ \ln \frac{2 \left(1-r\right) m_W^2 q^{-} v^{+}\omega}{\mu _F^4}-2\right]\right.
%-------------------------------
\\
%-------------------------------
\notag &+2\left(\ln^2\frac{m_b}{\mu _F}- \ln \frac{m_b}{\mu _R}\right)+2\mathrm{Li}_2(r)+\ln^2(1-r)
+\left[r+2 \ln \left(1-r\right)-1\right]\ln\frac{r}{1-r}
%-------------------------------
\\
%-------------------------------
&\left.+\frac{\pi ^2}{12}-i\pi \left(r+2\ln \frac{2 q^{-}
v^{+}\omega}{ m_W^2}-1\right)\right\},
%-------------------------------
\\
%-------------------------------
T_{\Sigma}^{(1)}(\omega) = &\frac{\alpha_s C_F}{4\pi}T^{(0)}(\omega)
\left(\ln\frac{2q^{-}v^{+}\omega}{\mu_R^2}-1-i\pi\right),
%-------------------------------
\end{align}
%-------------------------------
\label{T:one:bulk:of:diagrams}
%-------------------------------
\eseq
%-------------------------------
where we have introduced a dimensionless ratio for convenience:
%-------------------------------
\beq
%-------------------------------
r \equiv m^2_b/m^2_W.
%-------------------------------
\label{def:r:ratio}
%-------------------------------
\eeq
%-------------------------------

The LSZ reduction formula requests us also to consider the
wave function correction to the the $u$ quark,
which are represented by Fig.~\ref{Fig:One:loop:Amplitude}$f)$ and Fig.~\ref{Fig:One:loop:LCDA}$f)$.
They contribute to $T^{(1)}$ through
%-------------------------------
\begin{align}
%-------------------------------
\Phi_{+}^{(0)}\big(\omega\big)\otimes T_{\delta Z_u}^{(1)}
\big(\omega\big)= \mathcal{M}_{\delta Z_u}^{(1)} -
T^{(0)}\big(\omega\big)\otimes\Phi_{+\delta Z_u}^{(1)}\big(\omega\big)\,.
%-------------------------------
\label{T1:delta:Zu}
%-------------------------------
\end{align}
%-------------------------------
Obviously, the $u$ quark self-energy diagrams
are identical between Fig.~\ref{Fig:One:loop:Amplitude}$e)$ and Fig.~\ref{Fig:One:loop:LCDA}$e)$.
Substituting $\mathcal{M}_{\delta Z_u}^{(1)} =
\tfrac{1}{2}\delta Z_u(\mu_R)\mathcal{M}_{\delta Z_u}^{(0)}$ and $\Phi_{+\delta Z_u}^{(1)} =
\tfrac{1}{2}\delta Z_u(\mu_F)\Phi_{+}^{(0)}$ in \eqref{T1:delta:Zu}, one finds
%-------------------------------
\beq
%-------------------------------
T_{\delta Z_u}^{(1)}
(\omega)=\frac{1}{2}\left[\delta Z_u(\mu_R)-\delta Z_u(\mu_F)\right]T^{(0)}(\omega)
=\frac{\alpha_sC_F}{4\pi}\ln\frac{\mu_F}{\mu_R}T^{(0)}(\omega),
%-------------------------------
\label{T:one-loop:Zu}
%-------------------------------
\eeq
%-------------------------------
which simply vanishes once setting $\mu_R=\mu_F$.

In a similar vein, one needs also consider the external $\bar{b}$ leg correction
in QCD amplitude and HQET LCDA, as depicted in Fig.~\ref{Fig:One:loop:Amplitude}$f)$ and Fig.~\ref{Fig:One:loop:LCDA}$f)$.
The on-shell $b$-quark wave function renormalization constant in QCD reads
%-------------------------------
\beq
%-------------------------------
\delta Z_b\left(\mu\right)  = {\alpha_s C_F\over 4\pi}\left( -{1\over \epsilon_{\rm UV}}-
{2\over \epsilon_{\rm IR}} - 3 \ln {\mu^2 \over m_b^2} - 4\right).
%-------------------------------
\eeq
%-------------------------------

When computing the $\bar{b}$ quark self-energy contribution to $\Phi_{+}^{(1)}\otimes T^{(0)}$, one encounters a vanishing result
in DR. This is because the self-energy diagram of an on-shell $b$ quark in HQET yields a scaleless integral.
If we insist on distinguishing the UV and IR poles,
the $b$ quark on-shell wave function renormalization constant in HQET then reads
%-------------------------------
\beq
%-------------------------------
\delta \hat{Z}_{b}= {\alpha_s C_F\over 4 \pi}\left({2\over \epsilon_{\text{UV}}}-{2\over \epsilon_{\text{IR}}}\right),
%-------------------------------
\eeq
%------------------------------- .
which obviously has the same IR pole as $Z_b$ in full QCD.

Discarding the UV poles, the external $\bar{b}$ leg corrections yields the following
contribution to $T^{(1)}$:
%-------------------------------
\beq
%-------------------------------
T_{\delta Z_b}^{(1)} (\omega)=\frac{1}{2}\left[\delta  Z_b(\mu_R)-\delta  \hat
Z_b(\mu_F)\right]T^{(0)}(\omega)=\frac{\alpha_sC_F}{4\pi}
\left(2\ln\frac{m_b}{\mu_F}+\ln\frac{m_b}{\mu_R}-2\right)T^{(0)}(\omega).
%-------------------------------
\label{T:one-loop:Zb}
%-------------------------------
\eeq
%-------------------------------

In the leading power of $1/m_b$, it turns out that, reassuringly,
the box diagram in Fig.~\ref{Fig:One:loop:Amplitude}$c)$
makes vanishing contribution to the hard-scattering kernel.
Because the loop momentum generates the leading contribution to $\mathcal{M}^{(1)}_{\mathrm{box}}$
only in the {\it soft} region ($l^\mu \sim \Lambda_{\rm QCD}$),
which is already fully captured by $\Phi^{(1)}_{\mathrm{box}}\otimes T^{(0)}$. Therefore there arises no contribution to
$T^{(1)}_{\mathrm{box}}$, which exhibit the same pattern as what is observed
in the case of $B\to \gamma l\nu$~\cite{DescotesGenon:2002mw}.

Piecing all the relevant one-loop contributions in \eqref{T:one:bulk:of:diagrams},
\eqref{T:one-loop:Zu} and
\eqref{T:one-loop:Zb} together, we
can deduce the complete hard-scattering kernel at ${\cal O}(\alpha_s)$:
%-------------------------------
\begin{align}
%-------------------------------
\label{T1:full:expression}
%-------------------------------
T^{(1)}(\omega,m_b, \mu_F) =&\frac{\alpha_sC_F}{4\pi}\Bigg\{\ln
^2\frac{2 q^{-} v^{+}\omega}{\mu_F^2}-2
\ln^2\frac{m_{b}}{\mu_F}+\left(5-4 \ln\frac{1-r}{r}\right)
\ln\frac{m_{b}}{\mu_F}
%-------------------------------
\\
%-------------------------------
\notag &+2\mathrm{Li}_2(r)+\ln^2r-\left(2\ln\frac{1-r}{r}-3+r\right)
\ln\frac{1-r}{r}+\frac{\pi ^2}{12}-7
%-------------------------------
\nn\\
%-------------------------------
&- i\pi \left[2 \ln \frac{2 q^{-}
v^{+}\omega}{\mu_F^2}
-4\ln\frac{m_{b}}{\mu_F}-r-4\ln{\left(1-r\right)}+2\ln r+3\right]\Bigg\} T^{(0)}(\omega).\nn
%-------------------------------
\end{align}
%-------------------------------
It should be noticed that the unit mass $\mu_R$ has disappeared in the final answer,
as the consequence of the conserved (axial)vector current in QCD.
Nevertheless, $T^{(1)}$ still explicitly depends on
the factorization scale $\mu_F$. One can readily checks that, the
$\mu_F$ dependence of $T^{(1)}$ reads
%-------------------------------
\begin{align}
%-------------------------------
\mu_F\frac{d}{d\mu_F} T^{(1)}(\omega,\mu_F)=-\frac{\alpha_s C_F}{4\pi}\left(4\ln\frac{\omega}{\mu_F}+5\right)T^{(0)}(\omega)+\mathcal{O}\left( \alpha_s^2 \right).
%-------------------------------
\end{align}
%-------------------------------
Hearteningly, this specific $\mu_F$ dependence is exactly what we need.
In junction with the $\mu_F$ dependence of $\hat{f}_B(\mu_F)$ specified
in \eqref{hat:f:matching:fB} and the scale dependence of $\phi_B^+(\omega,\mu_F)$ governed by \eqref{LN:evolution:eq},
such a scale dependence of the hard-scattering kernel
guarantees that the physical amplitude $\mathcal{M}^{(1)}$
in \eqref{HQET:factorization:theorem} is independent of the artificial scale $\mu_F$.

We have formally assigned $m_W\sim m_b$ in order to justify the HQET factorization.
Nevertheless, we should admit that $m_W\gg m_b$ in the realistic world. The symptom of such
practical hierarchy is reflected in the large logarithm of $m_W/m_b$ in the hard-scattering kernel.
To see this more transparently, let us expand $ T^{(1)}(\omega,m_b, \mu_F)$ in \eqref{T1:full:expression}
to the zeroth order of $r$:
%-------------------------------
\begin{align}
%-------------------------------
\label{T1:omega:expanded}
%-------------------------------
\notag T^{(1)}(\omega,m_b, \mu_F) \Big\vert_{\rm expd}= &\frac{\alpha_s C_F}{4\pi}
\Bigg[\ln^{2}\frac{\omega}{\mu_F}-\ln^{2}\frac{m_b}{\mu_F}+\ln\frac{m_b}{\mu_F}
\left(5+2\ln\frac{\omega}{\mu_F}\right)
+\ln\frac{m_W^2}{m_b^2}\left(3+2\ln\frac{\omega}{m_b}\right)
%-------------------------------
\\
%-------------------------------
&+\frac{\pi^2}{12}-7-i\pi\left(3+2\ln\frac{\omega}{m_b}\right)\Bigg] T^{(0)}(\omega)+\mathcal{O}(r).
%-------------------------------
%-------------------------------
\end{align}
%-------------------------------
Interestingly, we notice that the occurrence of the single collinear logarithm ${\alpha_s C_F\over 4\pi}\ln{m_W\over m_b}$,
which might be large and potentially spoil the convergence of perturbative expansion.
We admit this is an inevitable drawback of the fixed-order calculation in the HQET factorization.
In the future work, we will illustrate how to apply the ERBL equation to effectively resum these types of
large logarithms through a refactorization program. The theoretical framework that renders such a
resummation program feasible is based upon a recently proposed factorization theorem
that links the LCDAs of $B$ meson defined in full QCD
and HQET~\cite{Ishaq:2019dst}.

With the ${\mathcal O}(\alpha_s)$ hard-scattering kernel at hand,
we are then able to present the NLO perturbative corrections to the form factors $F_{V/A}$:
%-------------------------------
\begin{align}
%-------------------------------
&F_V^{(1)}  = F_A^{(1)}=  F_{V/A}^{(0)} \int_0^\infty
\frac{d\omega}{\omega}
\frac{T^{(1)}\big(\omega\big)}{T^{(0)}\big(\omega\big)}
\phi_B^+\big(\omega\big)
%-------------------------------
\nn\\
%-------------------------------
& =  F_{V/A}^{(0)}  {\alpha_s C_F\over 4\pi}
\bigg\{-\ln^2{m_b\over \mu_F} -\ln{m_b\over \mu_F}
\left(2 \ln {1-r\over r}- 2\right)+ 2 {\rm Li}_2 (r)-\ln^2(1-r)
%-------------------------------
\nn\\
%-------------------------------
&+2\ln r\ln(1-r)+\left(3-r\right)\ln\frac{1-r}{r}+\frac{\pi^{2}}{12}-5-
2 \sigma_{B,1} \left(\ln\frac{1-r}{r}+\ln\frac{m_b}{\mu_F}\right)
%-------------------------------
\nn\\
%-------------------------------
&-\sigma_{B,2}+{i \pi}\left[2\ln\frac{m_b}{\mu_F}-3+r+2\ln\big(1-r\big)+2\sigma_{B,1}\right]\bigg\},
%-------------------------------
\label{eq:F_VA_01}
%-------------------------------
\end{align}
%-------------------------------
where the first inverse moment $\lambda_B^{-1}$ and the $n$-th logarithmic moment $\sigma_n\lambda_B^{-1}$ of the
$B$ meson LCDA have been defined in (\ref{first:inv:moment}) and (\ref{Log:first:inverse:moments}).
In the final expression, we have also utilized \eqref{hat:f:matching:fB} to
trade $\hat{f}_B$ in favor of the QCD decay constant $f_B$.
One can check that, the NLO prediction $F_{V/A}^{(0)}+F_{V/A}^{(1)}$ becomes independent of
the factorization scale $\mu_F$ up to the error at $\mathcal{O}(\alpha_s^2)$.

Equation~\eqref{eq:F_VA_01} constitutes the most important result of this work.
The equality between $F_V$ and $F_A$ at NLO in $\alpha_s$ looks peculiar, which
should not be simply viewed as a coincidence.
The underlying reason may be likely attributed to the heavy quark spin symmetry, which
is an exact symmetry at the lowest order in $1/m_b$.

It is instructive to compare \eqref{eq:F_VA_01}, which entails the NLO corrections rigorously
derived from HQET factorization, with the corresponding expressions of the form factors
for a similar process with $B^+$ meson replaced with the flavored quarkonium
$B_c^+$,  {\it e.g.}, $W^+\rightarrow B_{c}^+\gamma$~\cite{Feng:2019meh}.
This exclusive quarkonium production process has recently been investigated
through ${\cal O}(\alpha_s)$ within NRQCD factorization framework~\cite{Feng:2019meh}.
There the NRQCD short-distance coefficient for such a case contains three distinct scales:
$m_W$, $m_b$ and $m_c$. Besides the standard light-cone limit $m_W\gg m_b\sim m_c$,
the authors of \cite{Feng:2019meh} have also investigated the so-called ``heavy quark limit'',
where the NRQCD short-distance coefficient is expanded according to the scale hierarchy $m_W\sim m_b\gg m_c$.
In such a limit, the expanded NRQCD short-distance coefficient reads~\cite{Feng:2019meh}
%-------------------------------
\begin{align}
%-------------------------------
& F_{A}^{(1)}  = {F_{A}^{(0)}} \, {\alpha_s C_F\over 4\pi} \bigg\{
-\ln^2 x_0+  \ln x_0 \left(2 \ln \frac{1-r}{r}-5\right) + 2\mathrm{Li}_2(r) - \ln ^{2}(1-r)
%-------------------------------
\nn\\
%-------------------------------
& + 2 \ln(1-r)\ln r  + (3-r) \ln \frac{1-r}{r} -{2\pi^2\over 3}-9  + i \pi\left[ -2\ln x_0 -3+r+ 2\ln (1-r) \right]\bigg\},
%-------------------------------
\label{Bc:NRQCD:short-distance-coeff:Heavy-Quark-Limit}
%-------------------------------
\end{align}
%-------------------------------
where $x_0\equiv \tfrac{m_c}{m_b+m_c}\approx \tfrac{m_c}{m_b}$. It is amazing that  \eqref{Bc:NRQCD:short-distance-coeff:Heavy-Quark-Limit} resembles
\eqref{eq:F_VA_01} in many aspects, once the substitution $\mu_F = \omega \rightarrow m_c$ is
made in \eqref{eq:F_VA_01}. A simplification of this scale-fixing is that
the logarithmic moments $\sigma_{B,n}$ ($n=1,2$) drops out in \eqref{eq:F_VA_01}.
It looks peculiar that these two expressions  agree on the bulk
of (di-)logarithmic functions of $r$, but only differ
in constant~\footnote{To ``perfectly'' match
\eqref{Bc:NRQCD:short-distance-coeff:Heavy-Quark-Limit} with \eqref{eq:F_VA_01},
one may attempt to substitute the analytic $B^+$ meson LCDA defined in HQET~\cite{Bell:2008er},
where the $B^+$ meson is modelled as a free nonrelativistic $\bar{b} u$ pair.
The $\mu_F$ dependence in \eqref{eq:F_VA_01} would cancel explicitly,
after adding two convolution integrals $\Phi_{[\bar b u]}^{+(0)}\otimes T^{(1)}$ and
$\Phi_{[\bar b u]}^{+(1)}\otimes T^{(0)}$.
After the substitutions $\lambda_B\rightarrow m_c$, $\sigma_{B,1}\rightarrow -\ln m_c /\mu_F$ and $\sigma_{B,1}\rightarrow -\ln^2 m_c /\mu_F$ are made, \eqref{eq:F_VA_01} becomes almost identical to
\eqref{Bc:NRQCD:short-distance-coeff:Heavy-Quark-Limit}.}.

Similar to \eqref{T1:omega:expanded}, we can also expand the form factors $F_{V/A}^{(1)}$ to
the zeroth order in $r$:
%-------------------------------
\begin{align}
%-------------------------------
F_{V/A}^{(1)}\Big\vert_{\rm expd} = & {F_{V/A}^{(0)}} \frac{\alpha_s C_F}{4\pi}
\Bigg[\ln{m^2_b\over m^2_W}\left(2\ln{m_b \over \mu_F}+2\sigma_{B,1}-3\right)
-\ln^{2}\frac{m_b}{\mu_F}+2\left(1-\sigma_{B,1}\right)\ln\frac{m_b}{\mu_F}
%-------------------------------
\nn\\
%-------------------------------
&-\sigma_{B,2}+\frac{\pi^2}{12}-5+i\pi\left(2\ln\frac{m_b}{\mu_F}
+2\sigma_{B,1}-3\right)\Bigg]+\mathcal{O}(r).
%-------------------------------
\label{Form:factors:expanded:r}
%-------------------------------
\end{align}
%-------------------------------
Given the huge hierarchy between $m_W$ and $m_b$, we expect that this approximate
expression should be numerically quite close to the exact one in \eqref{eq:F_VA_01}.

%%%%%%%%%%%%%%%%%%%%%%%%%%%%%%%%%%%%%%%
\section{Numerical results}
%%%%%%%%%%%%%%%%%%%%%%%%%%%%%%%%%%%%%%%
\label{sec:phenomenology}
%%%%%%%%%%%%%%%%%%%%%%%%%%%%%%%%%%%%%%%

\begin{figure}[!htb]
\centering
\includegraphics[width=\textwidth]{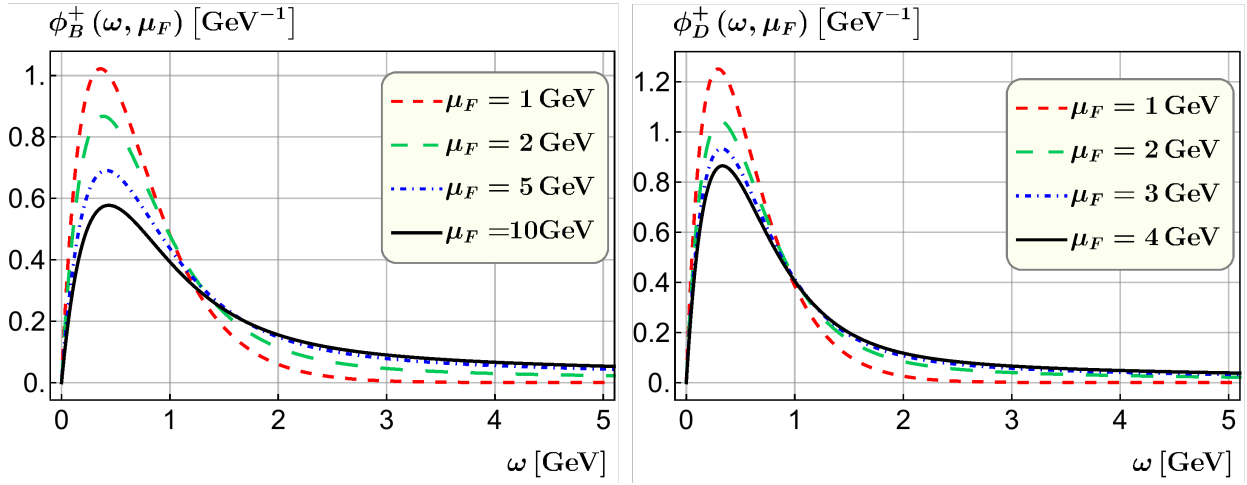}
\caption{Profiles of the LCDAs $\phi^{+}_{B/D_s}(\omega,\mu_F)$ at some typical values of the
factorization scale.} \label{fg:LCDA_Evltn}
\end{figure}

\begin{figure}[!htb]
\centering
\includegraphics[width=\textwidth]{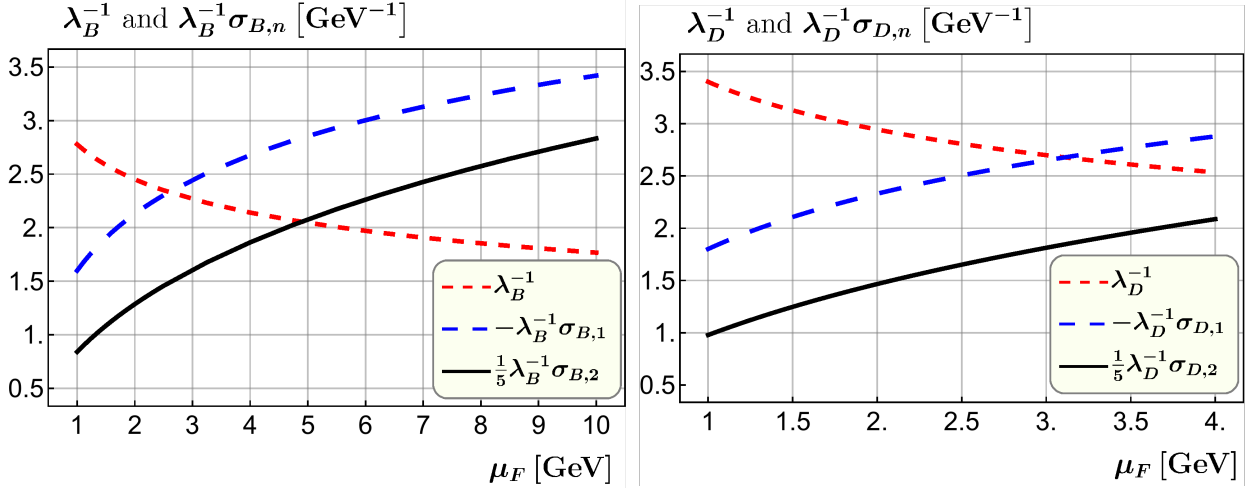}
\caption{Scale dependence of the first inverse moments $\lambda^{-1}_{M}$ and
the logarithmic inverse moments $\lambda^{-1}_{M} \sigma_{M,1/2}$ for $M= B^+, D_s^+$.
The renormalization scale ranges from 1 GeV to twice meson mass.}\label{Fig:Inverse:Moments:evol}
\end{figure}

In this section, we carry out detailed numerical predictions for vector/axial-vector form factors related to $W^+$
radiative decay into $B^+$ and $D_s^+$ mesons, as well as the corresponding partial widths and branching fractions.
The impact of the NLO perturbative corrections is also investigated.

We specify various input parameters as follows~\cite{Tanabashi:2018oca,Jegerlehner:2011mw}:
%--------------------------
\begin{table}[H]
\begin{centering}
\begin{tabular}{llll}
$\sin\theta_{W}=0.481,\;\;$ & $\alpha\left(m_{W}/2\right)=1/130,\;\;$ & $m_{W}=80.379\;\mathrm{GeV},\;\;$ & $f_{B}=0.187\;\mathrm{GeV},$ \tabularnewline
 $\left|V_{ub}\right|=3.65\times10^{-3},\;\;$ & $m_{b}=4.6\;\mathrm{GeV},\;\;$ & $m_{B}=5.279\;\mathrm{GeV},\;\;$& $f_{D}=0.249\;\mathrm{GeV},$\tabularnewline
$\left|V_{cs}\right|=0.997,\;\;$ & $m_{c}=1.4\;\mathrm{GeV},\;\;$ & $m_{D}=1.968\;\mathrm{GeV}.$\tabularnewline
\end{tabular}
\par\end{centering}
\end{table}
%--------------------------
We utilize the automated package
\texttt{HOPPET}~\cite{Salam:2008qg} to evaluate the QCD running coupling $\alpha_s$ with one-loop accuracy,
which has appropriately incorporated the effect associated with crossing the flavor threshold.

For the LCDAs of the $B^+$ and $D_s^+$ mesons defined at the initial scale $\mu_{F\,0}=1$ GeV,
we employ the simple exponential ansatz, first introduced by Grozin and Neubert~\cite{Grozin:1996pq}:
%--------------------------
\begin{align}
%--------------------------
\phi_{M}^{+}\big(\omega\big)=&\frac{\omega}{\lambda_M^2}\exp\left({-\frac{\omega}{\lambda_M}}\right),
%--------------------------
\end{align}
%--------------------------
with $\lambda_B=0.360$ GeV~\cite{Gelb:2018end} and $\lambda_{D_s}=0.294$ GeV~\cite{Yang:2016wtm}.

We would like to fathom out how the predicted decay rates depend on
the artificial factorization scale $\mu_F$.
To this purpose, we first need know how the LCDAs, and the corresponding (lograithmic) inverse moments, vary
with $\mu_F$.
The analytic solutions of the Lange-Neubert evolution equation in \eqref{LN:evolution:eq} have been
recently available~\cite{Lee:2005gza,Bell:2013tfa}.
Nevertheless in this work, we are content with numerically solving the Lange-Neubert equation via the
fourth order Runge-Kutta method, with the help of the package \texttt{GNU Scientific Library}~\cite{GSL}.
The profiles of the LCDAs for $B$ and $D_s$ at several factorization scales
are depicted in Fig.~\ref{fg:LCDA_Evltn}.

In Fig.~\ref{Fig:Inverse:Moments:evol}, we plot the scale dependence of the first inverse moment
and logarithmic inverse moments for both $B^+$ and $D^+_s$ mesons. We observe that the moments $\lambda_B^{-1}$
and $\lambda^{-1}_{M} \sigma_{M,1}$ exhibit a mild $\mu_F$ dependence between 1 GeV and twice meson mass,
whereas the inverse logarithmic moment $\lambda^{-1}_{M} \sigma_{M,2}$ develops a relatively
stronger $\mu_F$ dependence.

\begin{figure}[!htb]
\centering
\includegraphics[width=\textwidth]{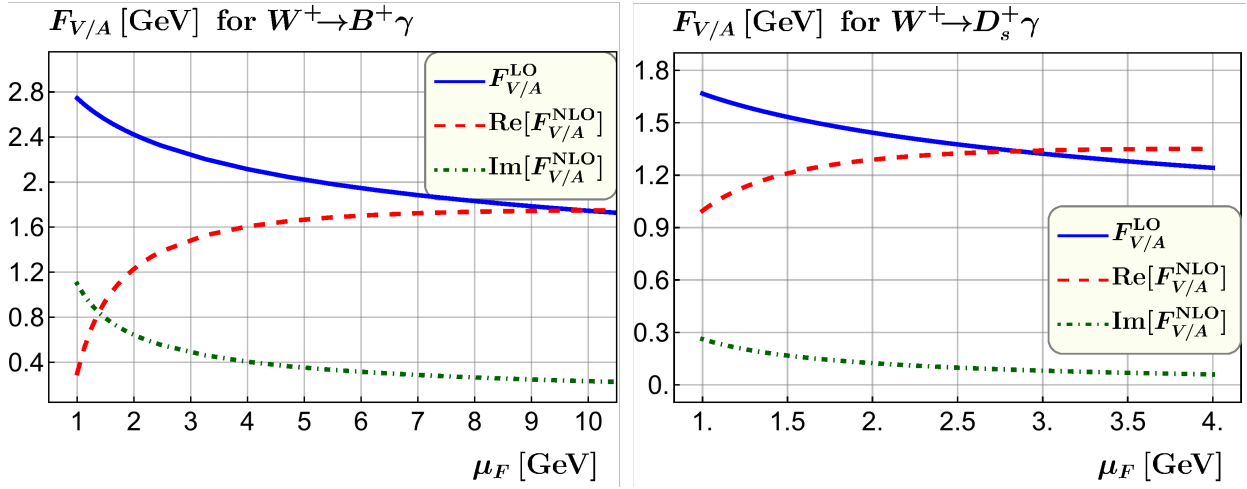}
\caption{Factorization scale dependence of the vector/axial-vector form factors $F_{V/A}$ at LO and NLO in $\alpha_s$,
for the process $W^+\to B^+\gamma$ and
$W^+\to D^+_s+\gamma$, respectively. The range of $\mu_F$ lies between 1 GeV to twice meson mass.}
\label{Fig:FromFactor:scale:dep}
\end{figure}

We define the LO and NLO predictions to form factors as
$F^{\rm LO}_{V/A}\equiv F^{(0)}_{V/A}$, and $F^{\rm NLO}_{V/A}\equiv F^{(0)}_{V/A}+F^{(1)}_{V/A}$,
At a given $\mu_F$, we evaluate the form factors through ${\cal O}(\alpha_s)$
in accordance with \eqref{FVA:LO:expression} and \eqref{eq:F_VA_01}.
In Fig.~\ref{Fig:FromFactor:scale:dep}, we plot the variation of the form factors with $\mu_F$,
at both LO and NLO accuracy. Clearly ${\mathcal O}(\alpha_s)$ correction is negative and significant for
both $W^+\to B^+\gamma$ and $W^+\to D_s^+\gamma$, especially with relatively small $\mu_F$.
Notice the scale dependence of the form factors becomes significantly reduced after incorporating the ${\mathcal O}(\alpha_s)$ correction, in particular with relatively greater $\mu_F$.

One can analytically prove that $F^{\mathrm{NLO}}_{V/A}$ in \eqref{eq:F_VA_01} is independent of
$\mu_F$ through $\mathcal{O}(\alpha_s)$. Notwithstanding considerable reduction of the $\mu_F$ dependence relative to the
LO prediction,  $F^{\mathrm{NLO}}_{V/A}$ in Fig.~\ref{Fig:FromFactor:scale:dep} still bear notable
factorization scale dependence, particularly in small $\mu_F$.
The residual scale dependence is clearly caused by neglected higher-order corrections.
In small $\mu_F$, the scale dependence may be amplified either by the large $\alpha_s$ or the large prefactors
accompanying $\ln \mu_F$.

\begin{figure}[!htb]
\centering
\includegraphics[width=\textwidth]{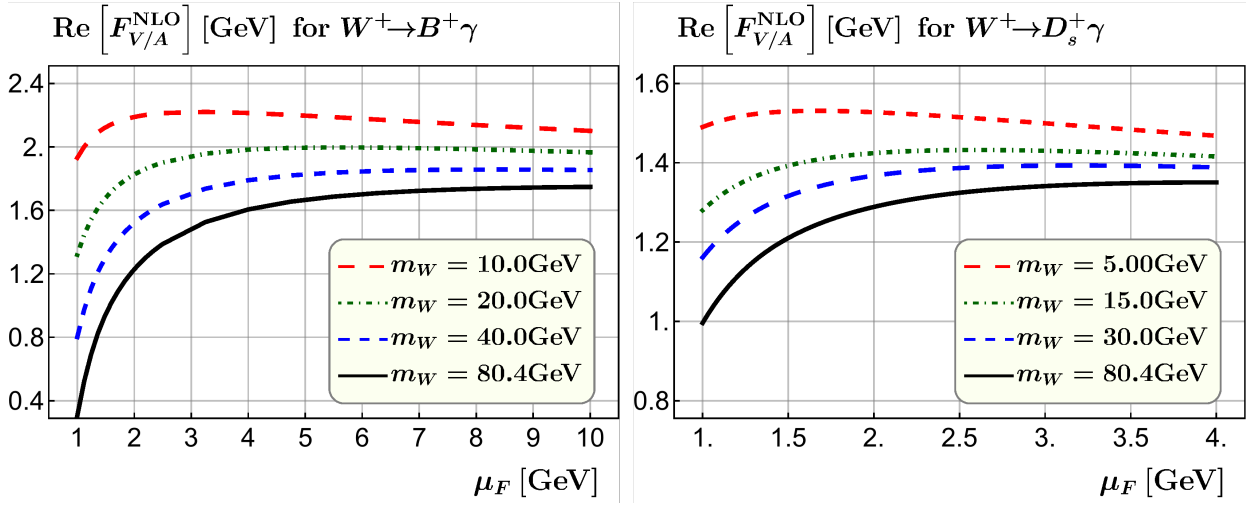}
\caption{Dependence of ${\rm Re}[F^{\rm NLO}_{V/A}]$ on $\mu_F$, chosen with some fictitious $W$ mass.
The factorization scale ranges from 1 GeV to twice meson mass.} \label{Fig:FromFactor:mW:dep}
\end{figure}

One source causing the residual $\mu_F$ dependence can be readily identified, that is,
the $\ln m^2_b/m^2_W \ln m_b/\mu_F$ term in \eqref{Form:factors:expanded:r}.
Numerically, the collinear logarithm is sizable, $\ln m_b^2/m_W^2\approx -5.7$,
which is largely responsible for the fall of the NLO prediction of the form factors with decreasing $\mu_F$.
In order to quantify our reasoning about the major source governing the residual scale dependence,
in Fig.~\ref{Fig:FromFactor:mW:dep} we artificially tune the value of $m_W$ from $80.38$ GeV down to $10$ GeV
for $W^+\to B^+\gamma$, and from physical mass down to $5\mathrm{GeV}$ for $W^+\to D^+_s\gamma$.
As can be clearly visualized from Fig.~\ref{Fig:FromFactor:mW:dep}, the smaller $m_W$ is chosen, the less scale dependence $F^{\mathrm{NLO}}_{V/A}$ does have in small $\mu_F$.

One might tend to conclude that, in order to reduce $\mu_F$ dependence in HQET factorization,
it appears desirable to resum the large collinear logarithm $\ln m_W^2/m_b^2$ appearing in the hard-scattering kernel to all orders
in $\alpha_s$. Note LN equation can only serve to sum large logarithm $\ln m_b/\Lambda_{\rm QCD}$, which has already included
in our numerical analysis. The occurrence of large logarithm $\ln m_W^2/m_b^2$ is a weakness of the
HQET factorization, since two distinct scales $m_W$ and $m_b$ has not been disentangled in this approach.
Resumming collinear logarithms of this type can only be accomplished by
appealing to the ERBL equation in the collinear factorization framework.
We recall that, for the analogous hard exclusive heavy quarkonium
production processes,  exemplified by $\gamma^*\to \eta_c+\gamma$ and $H\to J/\psi+\gamma$, the leading collinear logarithm of type $\ln Q/m_c$ has been identified and resummed to all orders in $\alpha_s$,
but the numerical effect turns out to be modest~\cite{Jia:2008ep}.
It is desirable if the similar goal can be achieved for exclusive heavy-light hadron production.

Very recently, we have proposed a novel factorization theorem, which attempts to
refactorize the $B$ meson LCDA in full QCD into the LCDA defined in HQET,
convoluted with a perturbatively calculable $Z$ function~\cite{Ishaq:2019dst}.
The underlying motivation is to separate the short-distance effect of order $m_b$ out of the QCD LCDA,
which cannot be a genuinely perturbative object.
Starting from the standard collinear factorization approach, armed with this factorization formula, we may
readily make an optimized prediction which combines the virtues of both collinear and HQET factorization approaches.
This improved approach will greatly facilitate resummation of both types of logarithms, $\ln m_b/\Lambda_{\rm QCD}$
and $\ln m_W^2/m_b^2$. We hope to present the optimized prediction for this process in future publication.

\begin{table}[!htb]
\begin{centering}
\begin{tabular}{|c|c|c|c|}
\hline
 & $\vphantom{\frac{L^L}{L^L}}\Gamma^{\mathrm{LO}}$ (GeV) & $\Gamma^{\mathrm{NLO}}$ (GeV)  & $\mathrm{Br}^{\mathrm{NLO}}$\tabularnewline
\hline
$\vphantom{\frac{L^L}{L^L}}W^{+}\rightarrow B^{+}\gamma$ & $\left(0.75\sim1.9\right)\times10^{-11}$ & $\left(3.1\sim7.7\right)\times10^{-12}$ & $\left(1.5\sim3.7\right)\times10^{-12}$\tabularnewline
$\vphantom{\frac{L^L}{L^L}}W^{+}\rightarrow D_{s}^{+}\gamma$ & $\left(0.72\sim1.3\right)\times10^{-7}$ & $\left(4.9\sim8.4\right)\times10^{-8}$ & $\left(2.3\sim4.0\right)\times10^{-8}$\tabularnewline
\hline
\end{tabular}
\par\end{centering}
\caption{Numerical predictions to the partial widths and branching ratios for the processes
$W^+\rightarrow B^+(D^+_s)\gamma$. The uncertainty is estimated by sliding $\mu_F$ from 1 GeV to twice meson mass.}
\label{Tab.nmrcl_prdctn}
\end{table}

\begin{figure}[!htb]
\centering
\includegraphics[width=\textwidth]{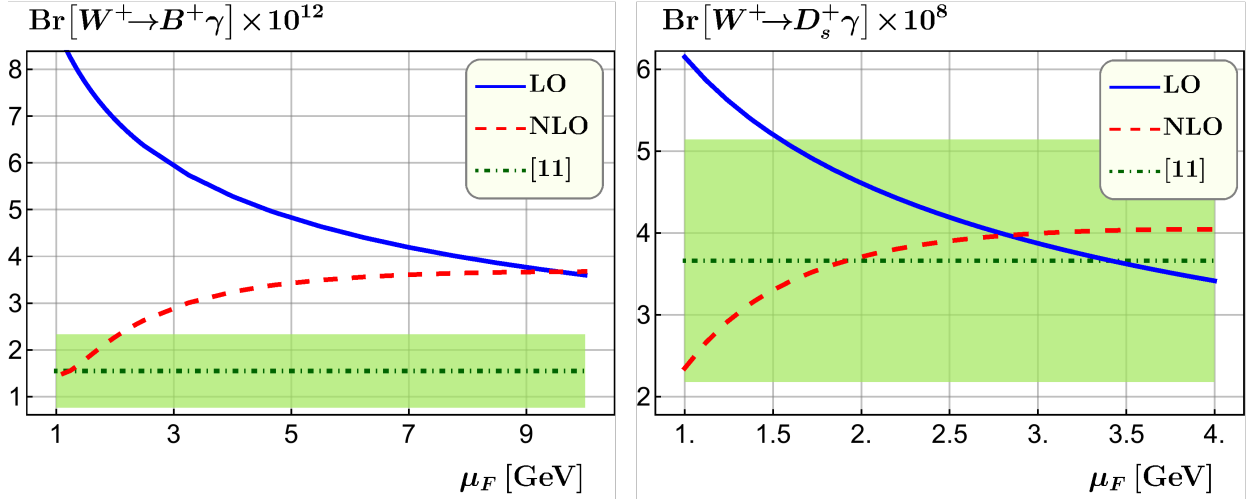}
\caption{Branching fractions of $W^+\rightarrow B^+\gamma$ and $W^+\rightarrow D_s^+\gamma$ as a function of
$\mu_F$, which ranges from 1 GeV to twice meson mass. Our predictions are juxtaposed with the existing ones obtained
from the collinear factorization~\cite{Grossmann:2015lea}, which are represented by the green bands. 
}\label{Fig:BR}
\end{figure}

We employ \eqref{eq:dcy_wdth} to compute the partial decay widths.
For the NLO prediction, we take the absolute square of the form factors, including their imaginary part in
$F_{V/A}^{(1)}$, without truncating the partial width strictly at ${\cal O}(\alpha_s)$.
We present the LO and NLO predictions for the partial widths and branching ratios of $W^+\to B^+\gamma$ and $W^+\to D_s^+\gamma$
in Table~\ref{Tab.nmrcl_prdctn} and Fig.~\ref{Fig:BR}, respectively.
Several orders of magnitude difference between $W^+\rightarrow B^+\gamma$ and $W^+\rightarrow D_s^+\gamma$ is primarily due
to $|V_{cs}|\gg |V_{ub}|$.

The NLO corrections turns out to be sizable, shift the LO prediction by $-83\%-2\%$ for $W^+\rightarrow B^+\gamma$, and
$-62\%\sim18\%$  $W^+\rightarrow D^+_s\gamma$, respectively.
Recall for a similar decay process $W\to B_c+\gamma$, the ${\cal O}(\alpha_s)$ correction has also been found to
considerably reduce the LO result~\cite{Feng:2019meh}.

Our state-of-the-art predictions for the branching fraction for $W^+\rightarrow B^+\gamma$ lies between
$(1.5-3.7)\times 10^{-12}$. It is difficult to observe such an extremely rare decay channel in the foreseeable future,
even after the integrated luminosity of LHC reaches 3000 ${\rm fb}^{-1}$.
On the other hand, the branching fraction for $W^+\rightarrow D_s^+\gamma$ is predicted to be within
$(2.3-4.0)\times 10^{-8}$, which may have bright observation prospect in the future LHC experiments.
Lastly, it is also interesting to remark that, our NLO predictions in HQET factorization are
somewhat greater than what were obtained from the standard collinear factorization approach for $W^+\rightarrow B^+\gamma$,
but compatible with theirs for $W^+\rightarrow D_s^+\gamma$,
albeit within large errors~\cite{Grossmann:2015lea}.

%%%%%%%%%%%%%%%%%%%%%%%%%%%%%%%%%%%%%%%
\section{Summary and outlook}
%%%%%%%%%%%%%%%%%%%%%%%%%%%%%%%%%%%%%%%
\label{sec:summary}
%%%%%%%%%%%%%%%%%%%%%%%%%%%%%%%%%%%%%%%

In this work, inspired by the NRQCD factorization for hard exclusive heavy quarkonium production,
we have formulated the HQET factorization approach, tailored for describing the
hard exclusive production of heavy-flavor meson.  This approach in spirit is quite different from
the standard collinear factorization for hard exclusive production.
We have taken the $W^+\rightarrow B^+(D^+_s)+\gamma$ as the prototype processes to
illustrate our theoretical framework, especially including the complete
NLO perturbative correction.
By examining that the NLO hard-scattering kernel is IR finite,
we have explicitly verified that the HQET factorization formula in \eqref{HQET:factorization:theorem} indeed
holds at first nontrivial order in $\alpha_s$ yet at lowest order in $1/m_b$.
It is conceivable that this factorization formula may hold to all orders in $\alpha_s$.

Interestingly, both vector/axial-vector form factors $F_{V/A}$ remain identical through NLO in $\alpha_s$, which may
be attributed to the the heavy-quark spin symmetry.
The NLO perturbative corrections turn to be substantial and negative. Our predictions for
$W^+\rightarrow B^+(D_s^+)+\gamma$ are compared with the existing ones using the collinear factorization approach.
While the $W^+\rightarrow B^+ +\gamma$ process is perhaps too suppressed to experimentally tag, the process
$W^+\rightarrow D_s^+ +\gamma$ may have positive chance to be observed in future LHC experiment.

A nuisance of the fixed-order calculation in HQET factorization approach is that the hard-scattering kernel
is inevitably plagued with large collinear logarithms of $m_W/m_b$, which may potentially ruin the
convergence of perturbative expansion. This reflects two hard scales $m_W$ and $m_b$ are not yet
disentangled in this approach.
Recently a new factorization theorem that links the $B$ meson
LCDAs defined between QCD and HQET has been discovered~\cite{Jia:2008ep}.
This factorization formula opens the gate for effectively merging both HQET factorization and collinear factorization
to make the optimized predictions. Within this new scheme, both large logarithms of
type $\ln m_b/\Lambda_{\rm QCD}$ and $\ln m_W/m_b$ can be efficiently resummed with the aid of LN equation and
ERBL equation. We hope to illustrate this improved theoretical framework in the future publication.

\begin{acknowledgments}
%-----------------------------------
We thank Feng Feng and Wen-Long Sang for valuable discussions. The Feynman diagrams in this manuscript
were prepared by using JaxoDraw~\cite{Binosi:2003yf}.
%-----------------------------------
The work of S.~I. and Y.~J. is supported in part by the National Natural Science Foundation of China under Grants No.~11875263,
No.~11621131001 (CRC110 by DFG and NSFC). S.~I. also wishes to acknowledge the financial support
from the CAS-TWAS President's Fellowship Programme.
%-----------------------------------
The work of X.-N.~X. is supported in part by the Deutsche Forschungsgemeinschaft (Sino-German CRC 110).
%-----------------------------------
The work of D.-S.~Y. is supported in part by the National Natural Science Foundation of China under Grants No.~11275263 and 11635009.
%-----------------------------------
\end{acknowledgments}

%%%%%%%%%%%%%%%%%%%%%%%%%%%%%%HQET LCDAs%%%%%%%%%%%%%%%%%%%%%%%%%%%%%%%%%%%%%%%%%%%%%%

\end{document}